\documentclass{aa}
\usepackage{txfonts}
\usepackage{graphicx}
\usepackage{natbib}
\usepackage[colorlinks]{hyperref}
\usepackage{booktabs}
\hypersetup{
    citecolor=[RGB]{44, 117, 255},  
    linkcolor=[RGB]{231, 62, 1},
    urlcolor=[rgb]{0,0,1}
}

\graphicspath{{./images/}}

\begin{document}

\title{Shape and connectivity of groups and clusters: \\ Effect of the dynamical state and accretion history}
\titlerunning{Shape and connectivity of groups and clusters}
\author{
  C.~Gouin\inst{1}\thanks{E-mail:~\tt{celine.gouin@ias.u-psud.fr}}, T. Bonnaire\inst{1,2}, N Aghanim\inst{1}
}
\institute{
Université Paris-Saclay, CNRS, Institut d'astrophysique spatiale, 91405, Orsay, France.
\label{inst1}
\and
Université Paris-Saclay, TAU team INRIA Saclay, CNRS, Laboratoire de recherche en informatique, 91190, Gif-sur-Yvette, France.
\label{inst2}
}
\date{\today}

\abstract
{
Matter distribution around clusters is highly anisotropic because clusters are the nodes of the cosmic web. The shape of the clusters and the number of filaments to which they are connected, that is, their connectivity, is thought to reflect their level of anisotropic matter distribution and must in principle be related to their physical properties.
We investigate the effect of the dynamical state and the formation history on both the morphology and local connectivity of about $2400$ groups and clusters of galaxies from the large hydrodynamical simulation IllustrisTNG at $z=0$.

We find that the mass of groups and clusters mainly affects the geometry of the matter distribution: Massive halos are significantly more elliptical and are more strongly connected to the cosmic web than low-mass halos. 
Beyond the mass-driven effect, ellipticity and connectivity are correlated and are imprints of the growth rate of groups and clusters. Both anisotropy measures appear to trace different dynamical states, such that unrelaxed groups and clusters are more elliptical and more connected than relaxed ones.
This relation between matter anisotropies and dynamical state is the sign of different accretion histories.
Relaxed groups and clusters have mostly been formed a long time ago and are slowly accreting matter at the present time. They are highly spherical and weakly connected to their environment, mostly because they had enough time to relax and thus lost the connection with their preferential directions of accretion and merging.
In contrast, late-formed unrelaxed objects are highly anisotropic with strong connectivities and ellipticities. These groups and clusters are in their formation phase and must be strongly affected by the infalling of materials from filaments.}


\authorrunning{Gouin et al.}

\maketitle

\section{Introduction}

Clusters of galaxies are not isolated structures, but rather the nodes of the large-scale cosmic web to which they are connected. Their birth and growth are intimately linked with the anisotropic nature of the gravitational collapse. Starting from nonspherical density peaks in the primordial density field \citep{Doroshkevich1970,Bardeen1986}, their asymmetries have been amplified by gravitation. Following the standard model of structure formation, clusters are also assumed to continuously grow by successive mergers and accretion along filamentary structures in the cosmic web \citep{Zeldovich1970}. The nonspherical shape of groups and clusters can therefore be used to test and validate the structure formation scenario \citep{Sereno2018} and the nature of dark matter \citep[DM,][]{Yoshida2000,Spergel2000,Dave2001,Peter2013}. 

On the one hand, the evidence of the nonsphericity of halos has been well established using cosmological N-body simulations \citep[see, e.g.,][]{Jing2002}. The elliptical shape of DM halos is expected to depend on their mass and redshift, as discussed in \cite{Allgood2006,Despali2014,Suto2016,Vega2017}, high-mass halos tend to be less spherical than low-mass objects.
This mass-ellipticity correlation can be attributed to the formation time \citep{Despali2014} and the formation history \citep{Chen2019,Lau2021} of halos. Following the hierarchical formation scenario, more massive structures formed more recently and are thus expected to be more widely disturbed because they did not have enough time to relax.
Investigating the link between the nonsphericity and the properties of groups and clusters is therefore an essential question for improving our understanding on the formation and growth of structures.
These investigations are also crucial to reduce uncertainties on the cosmological constraints. Any departure from spherical symmetry in the matter distribution can bias the cluster mass estimation \citep[see, e.g.,][]{Lee2018}, which in turn affects the cosmological constraints inferred from, for instance, the concentration mass relation \citep[see, e.g.,][]{Maccio2007}.

On the other hand, the large-scale environment of clusters is expected to affect their internal properties and their density profile \citep[see, e.g.,][]{Contigiani2020}. Much effort has recently been expended to probe relations between the brightest central galaxy (BCG) of groups and clusters with their local environment in the cosmic web \citep{Kuutma2020}. For example, \cite{Poudel2017} have probed the effect of cosmic web filaments on the properties of groups and their central galaxies. In addition, \cite{Darragh2019} have also investigated the relation of the number of filaments that are connected to groups and their BCG. The authors suggested that groups with high connectivity are more likely to have grown through a recent major merger, which tends to reduce the star-formation activity of the BCG. In general, the effect of cosmic web connectivity on the physical properties of galaxies has recently been investigated by \cite{Kraljic2020} and \cite{2019Lee}.

Moreover, the local topology and geometry of the density field next to massive halos can be probed statistically through the connectivity, that is, the number of filaments branching out from a halo. The statistics of halo connectivity is expected to depend on the growth factor, and thus can constitute an new topological constraint on dark energy \citep{Codis2018} because the connectivity of halos must decrease with cosmic time due to the accelerated expansion of the Universe that disconnects cosmic nodes from the network of filaments \citep{Pichon2010}. 
More generally, the anisotropic environment of the cosmic web affects the assembly bias of halos, as was theoretically explored by \cite{Musso2018}, who found that early-formed low-mass halos preferentially lie along filaments, while massive young halos are closer to the nodes \citep[see also][]{Shim2020,Cadiou2020}.
Cosmic web anisotropy is thus expected to be an indicator of halo assembly bias and is therefore strongly correlated with halo properties \citep[as discussed in][]{Paranjape2018,Ramakrishnan2019}.

In this context, galaxy clusters and groups are ideal places to describe the anisotropic level of matter distribution: from their nonspherical shape to the filaments that channel in-falling matter to clusters.
Recently, \cite{Gouin2020} have shown that the angular symmetries in the matter distribution increase with the distance from the cluster center. While matter inside clusters is well described by an ellipsoid (traced by quadrupolar moment), the filamentary pattern outside clusters show a complex azimuthal signature.
Both anisotropy signatures, traced by either the ellipticity or the connectivity, constitute probes for the structure formation scenario and cosmology. It is therefore of interest to understand the link between the mass assembly of clusters and the geometrical properties of their mass distribution.
In this study, we focus in particular on the three main questions: 
\begin{itemize}
    \item[$\bullet$] Does the anisotropic environment of groups and clusters influence their halo shape?
    \item[$\bullet$] Is the dynamical state of groups and clusters impacted by their anisotropic environment?
    \item[$\bullet$] Are the ellipticity and connectivity distributions resulting from different mass assembly histories?
\end{itemize}

To address these questions, we explore the shape of groups and clusters of galaxies and the large-scale filamentary structure that surrounds them using a large sample from the current cosmological hydrodynamical IllustrisTNG simulation \citep{ILLUSTRIS_TNG}.
The paper is organized as follows.
In Sect. 2 we define the sample of simulated groups and clusters and the quantities we used to describe the dynamical state and mass-assembly history of the halo.
In Sect. 3 we explain how we estimated the shape and the  connectivity of groups and clusters.
The statistics of ellipticity and connectivity are presented in Sect 4., where we explore possible correlations with the dynamical state and accretion history of the halo.
In Sect. 5 we discuss our findings in the context of recent studies of cluster evolution. 
Finally, Sect. 6 summarizes our main results.


\section{History and dynamical state of groups\label{SECT:data}}
\subsection{Groups and clusters from IllustrisTNG}

The hydrodynamical cosmological simulation IllustrisTNG \citep{ILLUSTRIS_TNG} follows the evolution of dark matter, gas, stars, and black holes on a moving mesh \citep[code Arepo,][]{AREPO} from redshift $z = 127$ to $z = 0$.
The input cosmology of the simulation is consistent with \textit{Planck} 2015 results \citep{Planck2016}, such that $\Omega_{\Lambda,0} = 0.6911$, $\Omega_{m,0} = 0.3089$, $\Omega_{b,0} = 0.0486$, $\sigma_{8} = 0.8159$, $n_s = 0.9667,$ and $h = 0.6774$.
We focus on the largest simulation box IllustrisTNG300-1 with a size length of $302.6 \text{ Mpc}$ and with the highest mass resolution of $m_{\text{DM}} = 4.0 \times 10^7 M_{\odot}/h$ ($2500^3$ DM particles).
We chose to use this high-resolution and large simulation box to accurately describe the shape and dynamical state of the massive halos. For each snapshot, the 
IllustrisTNG simulation provides a halo catalog, identified with a standard friends-of-friends (FoF) algorithm with linking length $b=0.2$ \citep{Davis1985}, and a subhalo sample derived with the Subfind algorithm to detect substructures inside host halos \citep{subfind}.
Based on the FoF group sample from IllustrisTNG, we selected $2522$ halos representing galaxy groups and clusters with masses $M_{200} \geq 1 \times 10^{13} M_{\odot}/h$ at $z=0$. We defined $M_{200}$ and $R_{200}$ as the radius of the group. $R_{200}$ is the radius enclosing $M_{200}$ with a mean overdensity of 200 times the critical background density. We discarded $79$ halos that are closer than $3 \, R_{200}$ from the simulation box edges to focus our exploration on the large-scale halo environments. 

\subsection{Dynamical state of groups and clusters \label{SECT:DS}}

In order to identify whether the anisotropies in matter distribution are correlated to the dynamical state, we categorized the groups and clusters by their level of relaxedness. The equilibrium state of each halo was assessed by computing three different parameters \citep[as defined in detail, e.g., by][]{Cui2017}.
First, we quantified the center-of-mass offset, which measures the distance between the center of mass $r_\text{cm}$ and the density peak of the halo $r_\text{c}$, normalized by the virial radius ($\Delta_\text{r} = \vert r_\text{cm} - r_\text{c} \vert / R_\text{vir}$). We considered $r_\text{c}$ to be the position at which the potential energy is minimum inside the halo, and $r_\text{cm}$ was computed as the sum of the mass-weighted relative coordinates of all particles and cells in the FoF halo.
Secondly, we determined the subhalo mass fraction which estimates the level of substructure in the halo. It is defined as $f_\text{sub} = \sum M_\text{sub} / M_\text{tot}$, where $M_\text{sub}$ is the mass of each subhalo without taking the mass of the main subhalo (i.e., the most massive subhalo) into account, and $M_\text{tot}$ is the total mass of the halo.
Thirdly, we computed the virial ratio of groups and clusters. It represents the level of halo virialization and is defined as $\eta = 2 T/\vert W \vert$, where $T$ and $W$ are the kinetic energy and the gravitational potential energy, respectively. It is computed from all particles and cells in the FoF halo.
 A cluster can thus be considered as relaxed when it has a low subhalo mass fraction ($f_\text{sub}<0.1$), a low center-of-mass offset ($\Delta_\text{r}<0.07$), and is close to virial equilibrium ($\vert \eta -1 \vert <0.15$). Halos that do not satisfy these three criteria can be classified as unrelaxed \citep[similarly to][]{Neto2007,Power2012}. In addition, we combined these three parameters to obtain a continuous measure of the level of the so-called relaxedness $\chi_\text{DS}$, as introduced by \cite{Haggar2020},
\begin{equation}
\label{EQ:relax}
    \chi_\text{DS} = \sqrt{ \frac{3}{ \left(\frac{\Delta_\text{r} }{0.07} \right)^2 + \left(\frac{f_\text{sub} }{0.1} \right)^2 + \left(\frac{\eta -1}{0.15} \right)^2 }}.
\end{equation}
Groups and clusters with $\chi_{DS} \geq 1$ were considered as dynamically relaxed \citep[see, e.g.,][]{Kuchner2020}.

\subsection{Mass accretion history and formation time \label{SECT:MAR}}

In order to probe the accretion history of groups and clusters, we estimated the instantaneous halo growth rate $dM/dt_{z\sim 0}$, the recent continuous mass accretion rate $\Gamma_{200\text{m}} (z=0.5)$, and the formation time $z_{\text{form}}$ of our sample. 
To do so, we used the merger tree of subhalos computed using the SubLink algorithm \citep[see][for details of the subhalo and merger tree computation in the Illustris simulation]{Rodriguez2015}. For each group, we focused on the merger tree of its main subhalo. Following the principal branch of the main subhalo merger tree, we extracted the mass ($M_{200\text{c}}$ and $M_{200\text{m}}$) of their associated halo at each time step.
We hence assume that the main subhalo progenitor is associated with the main halo progenitor.  This is almost always true, except in the occasional case of \textit{\textup{subhalo switching}} when a large number of particles from a given subhalo switch to a another subhalo between two consecutive snapshots \citep[this particular numerical case is fully described in][]{Poole2017}.

By computing the mass growth of halos over cosmic time, we can calculate different proxies of their mass-assembly history. 
Following \cite{Cole1996}, we defined the formation redshift $z_{\text{form}}$ as the redshift at which the mass of the halo main progenitor $M_{200}(z)$ is equal to half its mass at $z=0,$ such that
    \begin{equation}
    \frac{M_{200}(z_\text{form})}{M_{200}(z=0)} = 1/2.
    \end{equation}
The halo mass accretion controls the rate at which mass is deposited into halos. Previous works presented a large variety of definitions of the halo mass accretion rate. We focus on two of them. First, several studies describe the instantaneous mass accretion rate \citep[see, e.g., ][]{vandenBosch2002,Rodriguez2016} as
       \begin{equation}
            \frac{dM_{200\text{m}}}{dt}  = \frac{M_{200\text{m}}(t+dt)-M_{200\text{m}}(t)}{dt}.
        \end{equation}
       Because we are interested in quantifying the accretion of halo at the present time, we computed $dM/dt_{z=0.05}$ by performing a linear regression on $M_{200\text{m}}(t)$ over ten snapshots from $z=0$ to $z=0.11$. We note that $57$ low-mass halos have a negative recent growth $dM/dt_{z\sim 0} < 0$, probably due to \textit{\textup{particle switching}} \citep{Poole2017}. We discarded them for the specific investigations of the halo growth at $z \sim 0$.
Secondly, the logarithmic slope of the mass evolution $d \log M / d \log a $ is a natural description of the mass growth, following the parametric form of halo mass evolution over time $M(a) \propto \exp(-\alpha z)$, where $\alpha$ is related to the formation epoch of the redshift, and $a$ is the scale factor \citep[see, e.g.,][]{Wechsler2002}. This estimator of the continuous mass accretion history is defined as
        \begin{equation}
            \Gamma_{200\text{m}} = \frac{\Delta log(M_{200\text{m}})}{\Delta log(a)}.
        \end{equation}
 The halo mass changes through the physical accretion and the changes in the halo radius that are defined by the reference density. This definition of mass accretion rate $\Gamma$ thus accounts for the real physical accretion rate and has the advantage of isolating effects of pseudo-evolution in the cluster mass \citep{Diemer2013b}.
 Computing the accretion rate $\Gamma$ allows us to determine the accretion phase of halos between two times, from $z=0$ to $z\sim 0.5$. The redshift interval of $\Delta z =0.5$  corresponds reasonably well to the expected relaxation timescales of halos \citep{Power2012,Diemer2014,More2015,Chen2019}.
 By using the masses of halos at $z=0$ and their main progenitor at $z=0.5$, we thus computed the continuous accretion rate $\Gamma_{\text{200m}} (z=0.5)$ for each halo. $\text{Four}$ halos do not have a main progenitor at $z=0.5$ and were therefore discarded from the analyses for which the accretion history was used.


\section{Probing the anisotropy of matter distribution \label{SECT:METHOD}}

We aim to probe the anisotropic level of the density field in environments of massive objects.
The anisotropy of matter distribution was estimated at two different radial scales: inside groups and clusters, and in their environments.
 \cite{Gouin2017,Gouin2020} have recently reported that deviations from spherical symmetry increase from the cluster centers to their external regions with two distinct regimes: inside halos, and outside the virial radius. Inside clusters, the matter distribution is highly quadrupolar, reflecting the ellipsoidal shape of the halos. In cluster outskirts, from $1 \times R_\text{vir}$ to $4 \times R_\text{vir}$, the matter distribution describes more complex angular symmetries up to high harmonic orders. These high levels of angular harmonics are induced by the filamentary pattern around clusters and increase for clusters that connected to a large number of filaments \citep{Gouin2020}.
Following these previous investigations, we quantified the anisotropy inside groups and clusters by estimating their ellipticity, such that elongated shapes are defined as more anisotropic than spherical shapes. The anisotropy of the large-scale environment of groups and clusters was also estimated by counting the number of filaments that are connected to them. 
By extracting the cosmic filamentary structure, we considered as highly anisotropic environments those halo outskirts that are connected to a large number of preferentially accreting directions. 

\subsection{Ellipticity measurement}

The anisotropy of the matter distribution in halos was estimated by modeling the halo shape as an ellipsoid and by computing its ellipticity. The ellipticity is commonly defined as in \cite{Jing2002},
\begin{equation}
    \epsilon = \frac{\lambda_1 - \lambda_3}{2 \tau},
\end{equation}
where $\lambda_{\left[1,2,3\right]}$ are the axes of the ellipsoid with $\lambda_1 \geq \lambda_2 \geq \lambda_3$, and $\tau = \lambda_1 + \lambda_2 + \lambda_3 $. \\
For each group of clusters, we determined the major, intermediate, and minor axis vectors of the ellipsoid ($a>b>c$) by diagonalizing their mass tensor,
\begin{equation}
    I_{\alpha \beta} = \sum^N_{i=1} x^{(i)}_{\alpha} x^{(i)}_{\beta},
\end{equation}
where $x^{(i)}_\alpha$ ($\alpha = 1, 2, 3$) is the coordinate of the $i$th DM particle along the three axes.
Following \cite{Suto2016}, we used all the DM particles including all substructures and non-FOF members.
Starting from the sphere centered on the center of mass, we computed the eigenvalues of the mass tensor and fixed the size of the ellipsoid such that the total mass enclosed in the ellipsoid equaled $M_\text{ellipsoid}=M_{200}$. We then repeated the procedure until all the eigenvalues converged to within 1\%.
The axes of the ellipsoid were defined by the square roots of the tensor eigenvalues. The ellipsoidal shapes of halos for different values of the ellipticity are shown in Fig. \ref{FIG_VISU}.

\begin{figure}
    \centering
    \includegraphics[width=0.48\textwidth]{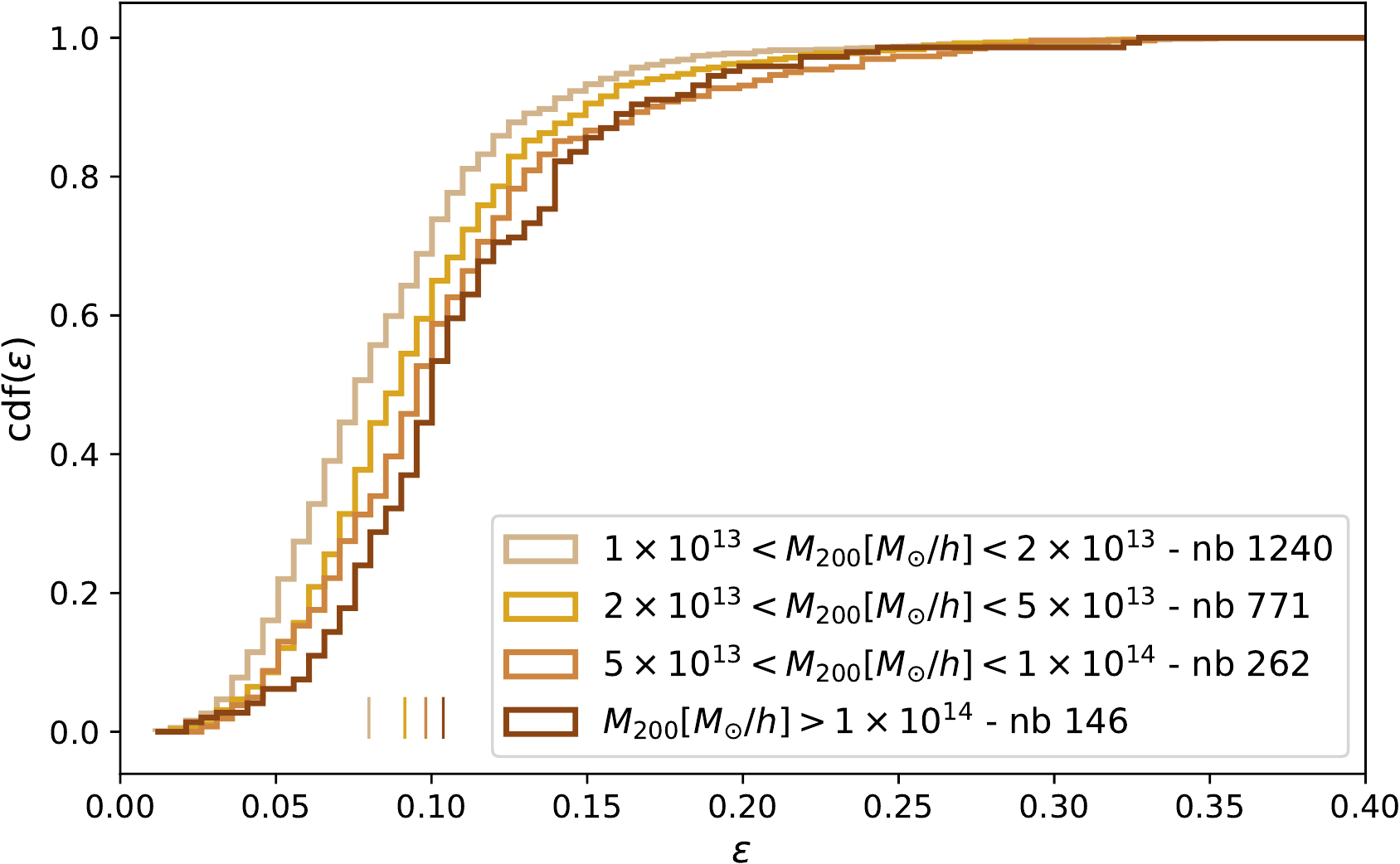}
    \caption{Ellipticity cumulative distribution for four different mass bins of groups and clusters at $z=0$. The short vertical lines at the bottom show the ellipticity median value for each mass bin. The number of groups and clusters per mass bin is written in the box.
    \label{fig:FIG_E_mass}}
\end{figure}

Figure \ref{fig:FIG_E_mass} displays the cumulative distribution function of the ellipticity for several mass bins. We confirm the results from previous theoretical studies that were also based on numerical simulations \citep[see, e.g.,][]{Allgood2006,Despali2014,Bonamigo2015,Vega2017,Despali2017} and show that massive groups and clusters on average have a more elliptical and thus a more anisotropic matter distribution.
This ellipticity-mass dependence can be attributed to the hierarchical formation scenario that implies that massive structures are formed more recently and thus have not had time to relax \citep{Despali2014}.

\subsection{Connectivity measurement}

The anisotropy in the large-scale environment of groups and clusters is quantified by the local number of filaments that are connected to them, the so-called connectivity $\kappa$. This proxy is a powerful tool for understanding the geometry of the underlying density field, as discussed in \cite{Codis2018}.
Different methods are proposed in the literature to extract the skeleton of the cosmic web \cite[for a review of recent methods, see][]{Libeskind2017}. We used the recently developed cosmic web detection algorithm T-ReX \citep{Bonnaire2020}, which builds an estimate of the filamentary pattern based on a graph modeling of the subhalo distribution. T-ReX extracts a smooth version of the minimum spanning tree using a Gaussian mixture model to describe the spatial distribution of tracers.
Two main parameters are used in the algorithm: one controls the smoothness of the graph, $\lambda,$ and one governs the level of the denoising pre-procedure, $l$. We considered here $\lambda=1$ and $l=5$ to capture the large-scale cosmic filaments and smaller bridges of matter.

Following \cite{DANY_2020a, DANY_2020b}, we applied the cosmic web detection algorithm on subhalos from the IllustrisTNG simulation selected according to their stellar masses, $M_{*} \, [M_{\odot}] \geq 10^9$. From the output graph, we associated the closest node of the graph with each group and cluster, and the local connectivity $\kappa$ is the number of intersecting filaments in a sphere of $1.5\, R_{200}$ radius around the graph node \citep[similarly to][]{Darragh2019,Sarron2019}.
We discarded 24 low-mass groups that were considered too distant from the graph ($>1$ Mpc/h). These objects are located in low-dense environments, and are thus not connected to the overall cosmic web. As a result, our final sample of groups and clusters consisted of $2419$ halos, for which we estimated the ellipticity and connectivity in order to probe the anisotropy inside and around halos.

\begin{figure}
    \centering
    \includegraphics[width=0.48\textwidth]{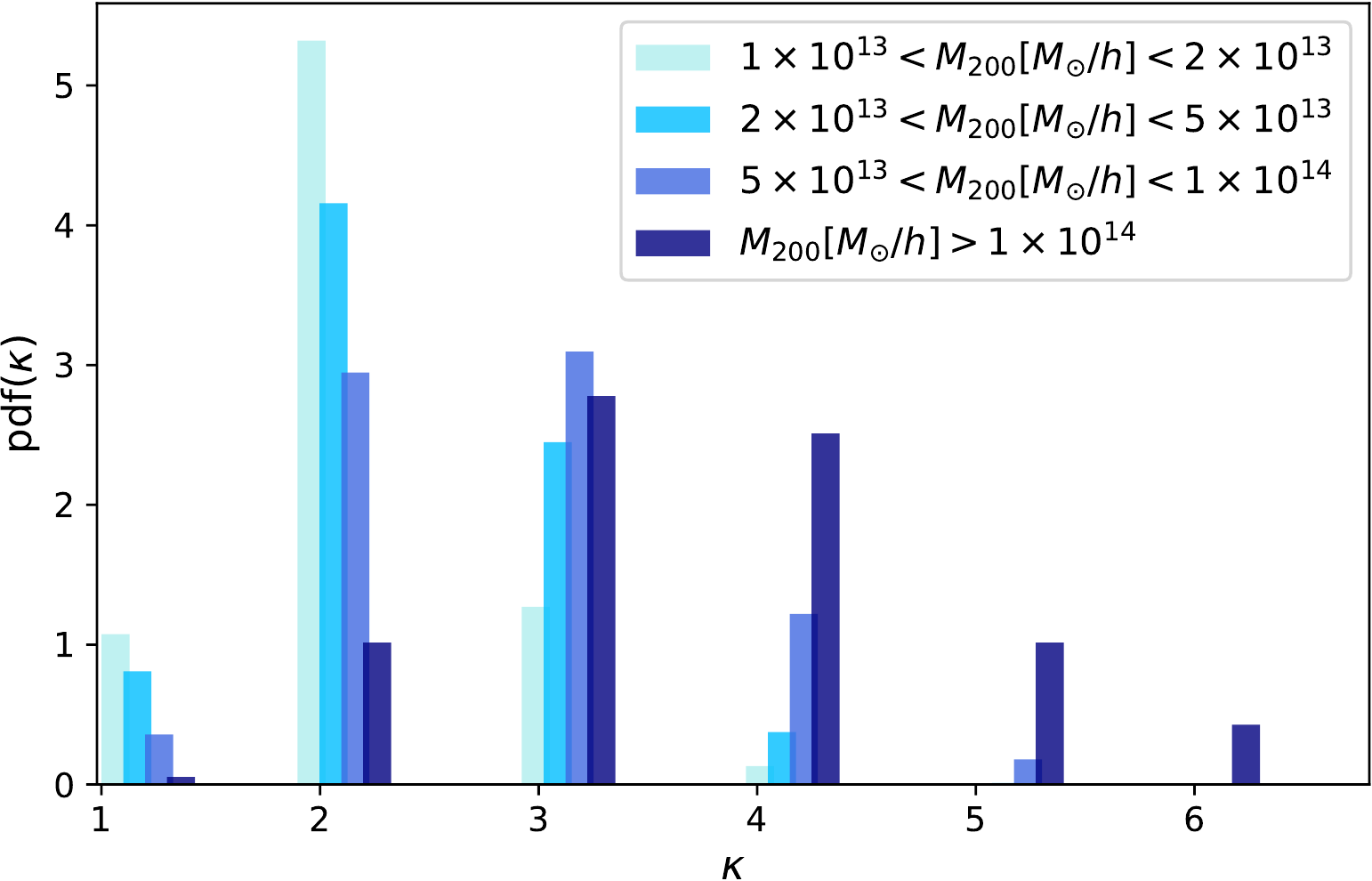}
    \caption{Probability distribution function of group and cluster connectivities $\kappa$ for four mass bins. Because $\kappa$ is an integer, we artificially shift the connectivity of the different mass bins by adding $+0.1$ for the second mass bin, $+0.2$ for the third mass bin, and $+0.3$ for the most massive mass bin for clarity.
    \label{fig:FIG_K_mass}}
\end{figure}

\begin{figure*}
    \centering
    \includegraphics[width=0.42\textwidth]{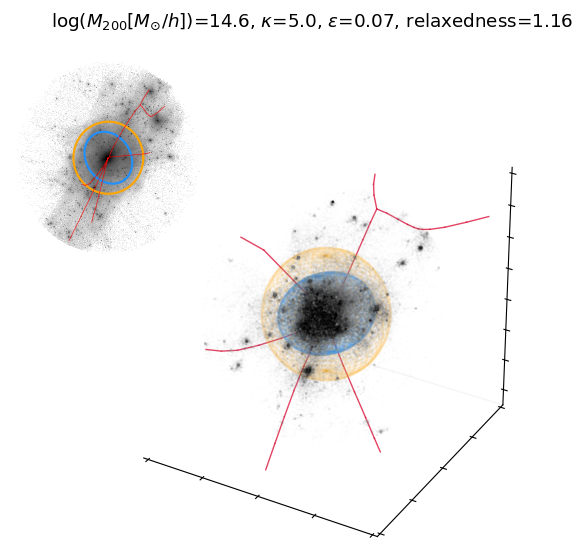} \includegraphics[width=0.42\textwidth]{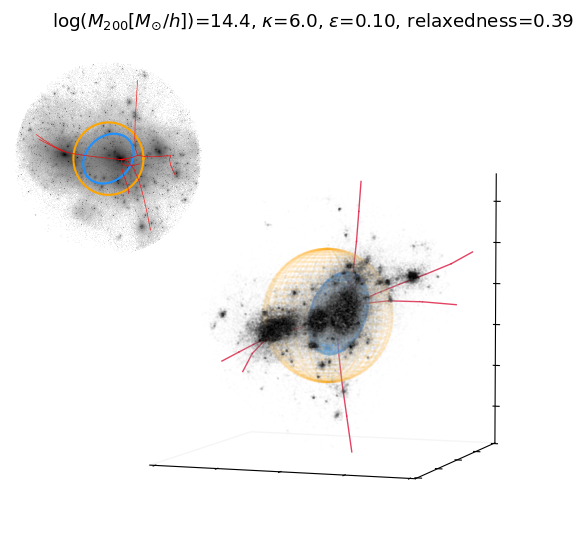} 

 \includegraphics[width=0.42\textwidth]{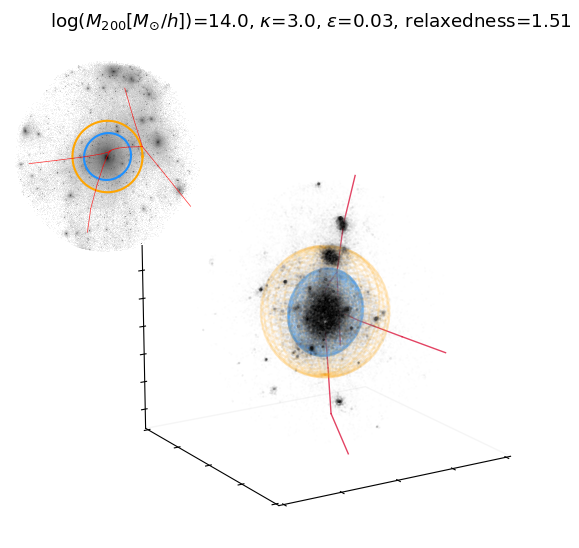} \includegraphics[width=0.42\textwidth]{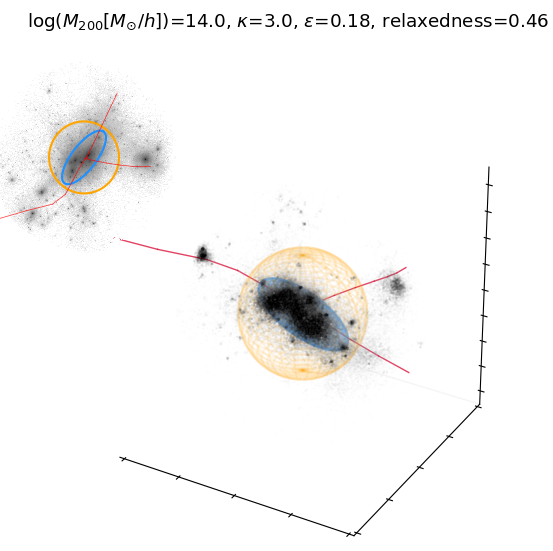}

    \caption{Ellipsoidal shapes and large-scale environments of four simulated galaxy clusters.
    The blue ellipsoids are computed by calculating the mass tensor of the DM distribution, following Sect. \ref{SECT:METHOD}. 
    The red lines represent cosmic filaments reconstructed by the T-ReX algorithm for the galaxy distribution. The yellow spheres have a radius of $1.5 \times R_{200}$. \label{FIG_VISU} }
\end{figure*}

The cluster connectivity is illustrated in Fig. \ref{FIG_VISU}. The T-ReX filamentary structure (computed from the galaxy distribution) clearly traces the DM distribution around halos well.
In addition, Fig. \ref{fig:FIG_K_mass} shows the probability distribution function of the connectivity for four mass bins. The histogram of the clusters ($M_{200} > 1\times 10^{14} M_{\odot}/h$) peaks around 3-4 (i.e., clusters are connected to three to four filaments) and is spread over high $\kappa$ values (up to $\kappa = 6$). In contrast, the connectivity statistics of the lowest-mass groups ($1\times 10^{13} <M_{200} [M_{\odot}/h] < 2\times 10^{13} $) strongly peaks at $\kappa=2$, suggesting that low-mass groups are preferentially located inside filaments, whereas massive clusters are more likely located at the nodes of the cosmic web. 
We thus confirm that massive structures have a higher connectivity than low-mass structures (as was also found by \citep[see, e.g.,][]{Aragon2010,Codis2018,Darragh2019,Sarron2019,Malavasi2020}).
In addition to the mass-connectivity relation, we explored any possible correlation between anisotropic environments (via the connectivity) and the properties of the groups and clusters.


\section{Results}

In this section, we assess the effect of the dynamical state and mass-accretion history on the anisotropies of the matter density field around simulated groups and clusters as traced by the ellipticity in their inner parts and the connectivity at larger scales where the groups and clusters are connected to the cosmic web.
Both anisotropy proxies depend on the mass: Massive structures are more elliptical and more strongly connected to the cosmic web than low-mass structures (see Figs. \ref{fig:FIG_E_mass} and \ref{fig:FIG_K_mass}). 
Our purpose is to determine whether there are any other correlations between the anisotropy and the halo properties beyond the mass-driven trend.
Therefore we adopted the strategy of exploring the relations of the physical halo properties and halo anisotropy proxies for different mass bins.
For all figures, we display the mean quantity together with their associated error bars derived from bootstrap re-sampling. To further assess the correlation between a set of parameters representing either the physical properties or the anisotropy proxies, we list the Spearman rank correlation coefficient $\rho_\text{sp}$ and the associated \textit{p}-value in Table \ref{tab:correl}. 

\begin{table*}[]
    \renewcommand{\arraystretch}{1.2}
    \centering
    \begin{tabular}{c|c c|c c}
      & \multicolumn{2}{c|}{$M_{200}>1 \times 10^{13} M_{\odot}/h$} & \multicolumn{2}{c}{$M_{200}>5 \times 10^{13} M_{\odot}/h$} \\ \cline{2-5}
      &  $\rho_\text{sp}$ & \textit{p}-value & $\rho_\text{sp}$ & \textit{p}-value \\ \specialrule{.12em}{.05em}{.05em} 
      
        $\epsilon$ and $\kappa$ & $0.11$ & $2.15 \times 10^{-7}$ & $0.17$ & $5 \times 10^{-4}$  \\
         \hline
         $\epsilon$ and  $dM/dt (z\sim 0)$ &  $0.37$ & $3.1 \times 10^{-78}$  & $0.31$ & $2 \times 10^{-10}$ \\
         $\kappa$ and $dM/dt (z\sim 0)$ &  $0.33$ & $3.8 \times 10^{-60}$  & $0.35$ & $4 \times 10^{-13}$ \\
         \hline
         $\epsilon$ and $\chi_\text{DS}$ &  $-0.44$ & $1.9 \times 10^{-7}$  & $-0.42$ & $1 \times 10^{-10}$ \\
         $\kappa$ and $\chi_\text{DS}$ &  $-0.13$ & $4.6 \times 10^{-11}$ & $-0.17$ & $5 \times 10^{-4}$ \\
    \end{tabular}
    \caption{Spearman rank correlation coefficients $\rho_\text{sp}$ and corresponding \textit{p}-values of halo properties and their anisotropy proxies for all the groups and clusters in the sample ($M_{200}>1\times 10^{13} M_{\odot}/h$), and for the 408 most massive groups and clusters ($M_{200}>5 \times 10^{13} M_{\odot}/h$). 
    \label{tab:correl}}
\end{table*}

\subsection{Effect of cosmic filament accretion on shapes}

\begin{figure}
    \centering
    \includegraphics[width=0.45\textwidth]{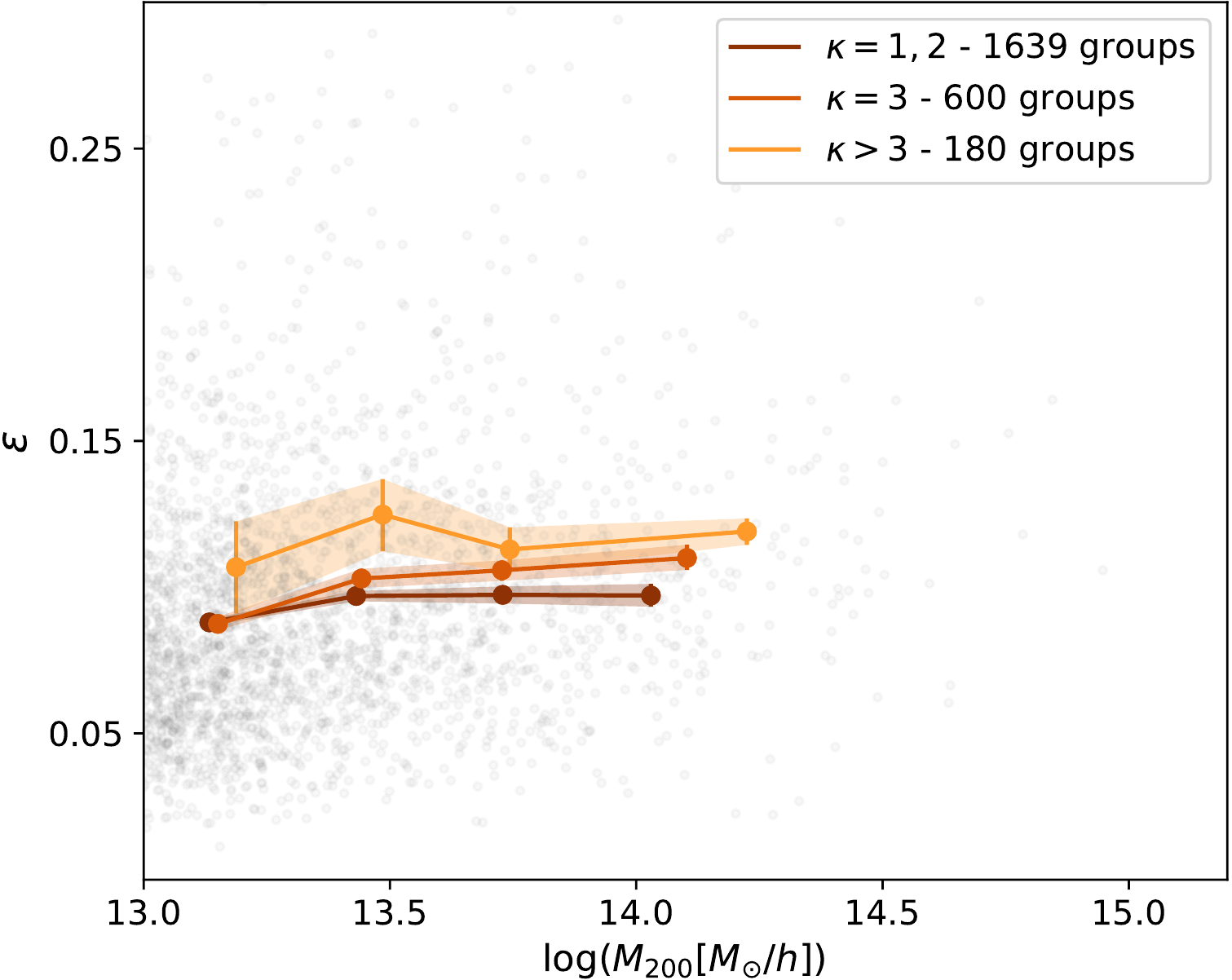}
    \caption{Mean ellipticity as a function of the mass for three connectivity bins: low ($\kappa \in \{1, 2\}$), medium ($\kappa =3$), or high ($\kappa > 3$).
    The error bars are the errors on the mean, derived from bootstrap resampling. In gray we show the point distribution of the sample.
\label{fig:FIG_E_K} }
\end{figure}

\begin{figure*}
    \centering
    \includegraphics[width=0.8\textwidth]{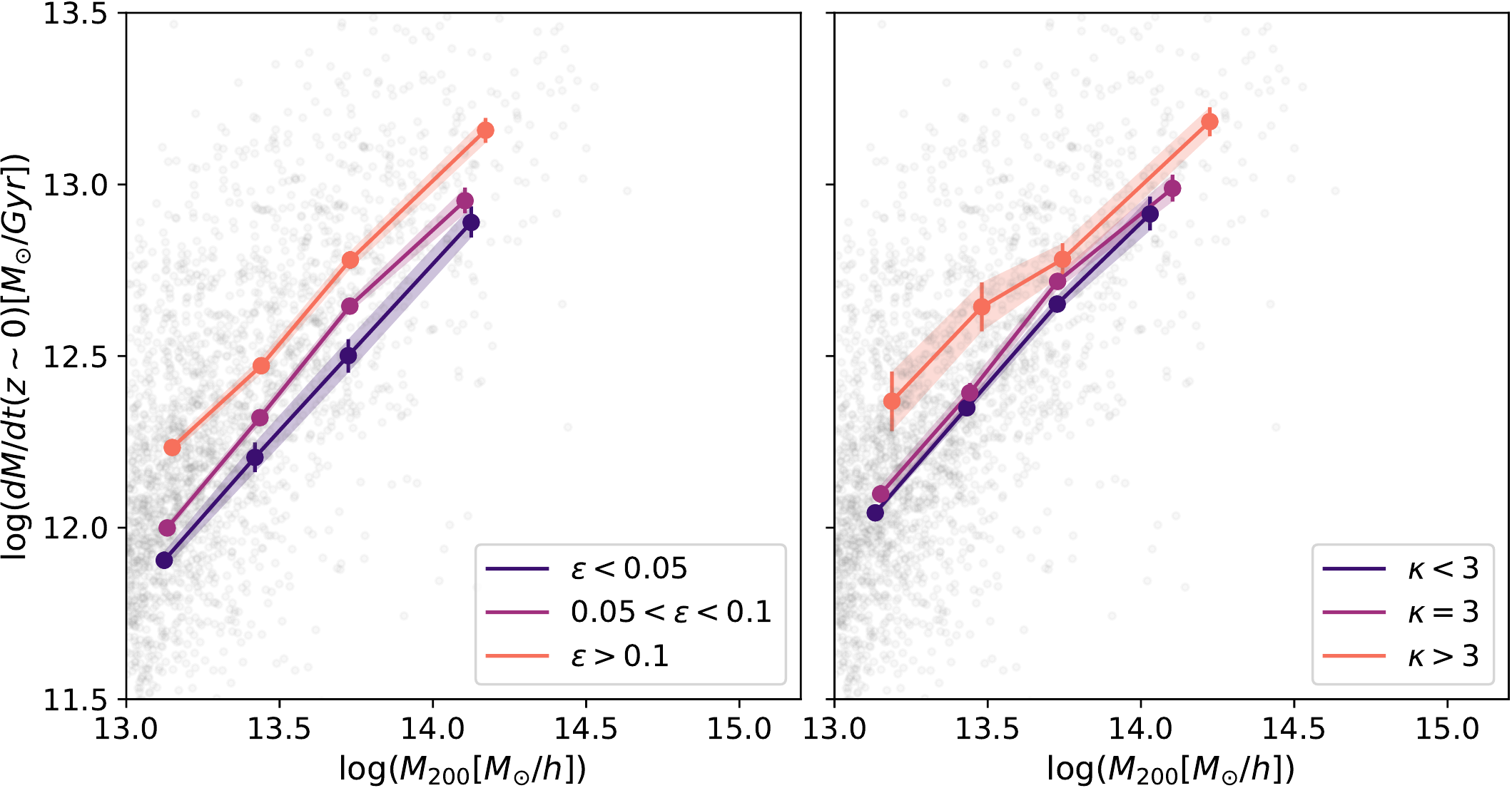} 
    \caption{Mean instantaneous mass growth $dM/dt$ computed at $z\simeq0$ as a function of the group mass for three ellipticity bin (left panel) and three connectivity bins (right panel).
    The ellipticity is binned according to an almost spherical ($\epsilon<0.05$), a quite elliptical ($0.05<\epsilon<0.1$), and an elliptical shape ($\epsilon>0.1$).
    The connectivity is binned such that clusters have a low ($\kappa \in \{1, 2\}$), medium ($\kappa =3$), or high ($\kappa > 3$) connectivity.
    The error bars are the errors on the mean, derived from bootstrap resampling. In gray we show the point distribution of the sample.
    \label{fig:dmdt} }
\end{figure*}

The effect of the anisotropic large-scale environment on the shape of groups and clusters is investigated first.
Given the known correlations of mass and connectivity, and mass and ellipticity, we aim at investigating whether connectivity and ellipticity are correlated beyond the mass effect.
Therefore Fig. \ref{fig:FIG_E_K} displays the mean halo ellipticity as a function of mass for three different connectivity bins.
We separately consider weakly connected ($\kappa = \{1, 2\}$), medium connected ($\kappa = 3$), and highly connected ($\kappa > 3$) groups and clusters.
At fixed mass, there is a trend for weakly connected objects to be more spherical than highly connected objects.
In contrast, structures with a high connectivity ($\kappa>3$) are on average more elliptical than the other structures.
This trend between connectivity and ellipticity is more significant for masses higher than $10^{13.5} M_{\odot}/h$, probably because the range of connectivities is wider for these mass bins.
The connectivity of low-mass groups ($10^{13}<M_{200}<10^{13.5} M_{\odot}/h$) is low, with about $80\%$ with $\kappa \in \{1,2\}$, suggesting that they are preferentially located inside filaments \citep[as shown by][]{Cautun2014, Hellwing2020}.
In contrast, most massive objects ($M_{200}>10^{14.2} M_{\odot}/h$) are largely connected to the cosmic web (about $70\%$ of them with a connectivity $\kappa>3$), showing that they are rather located at the nodes of the cosmic web.
Furthermore, as Table \ref{tab:correl} shows, the correlation coefficient between ellipticity and connectivity increases for high-mass halos and reaches $\rho_\text{sp} \sim 0.24$ for $M_{200} \geq 1 \times 10^{14} M_{\odot}/h$. This correlation remains significant for masses $M_{200} > 5 \times 10^{13} M_{\odot}/h$ when the partial correlation coefficient is computed, where the effect of mass dependence is corrected for using the fitting formulae of $\epsilon(M_{200})$ and $\kappa(M_{200})$.
We conclude that for masses higher than $10^{13.5} M_{\odot}/h$, the higher the connectivity, the more elliptical the cluster on average.

To understand this anisotropic environmental effect on the shape, we probed the possible relations between the accretion of infalling matter at the present time and the halo anisotropy proxies (both ellipticity and connectivity). In addition to the overall picture that groups and clusters are preferentially aligned with their main connected filaments as supported by previous theoretical \citep{Gouin2017,Okabe2020_SIMU,Kuchner2020} and observational \citep{Einasto2020,Okabe2020_OBS,Gouin2020} studies, \citet{Morinaga2020} have recently shown that ellipticity is correlated with the filamentary accretion.

The accretion of groups at $z=0$ can be quantified by the instantaneous mass accretion rate $dM/dt_{z \sim 0}$  (a proxy for the growth rate), as described in Sect. \ref{SECT:MAR}.
Fig. \ref{fig:dmdt} shows the mean instantaneous accretion rate of groups and clusters as a function of their masses. As expected following the hierarchical structure formation scenario, massive halos grow more rapidly on average than low-mass halos.
In addition to the mass-driven effect, the left panel of Fig. \ref{fig:dmdt} shows the instantaneous accretion rate for three bins of ellipticity: an almost spherical ($\epsilon<0.05$), a quasi-elliptical ($0.05<\epsilon<0.1$), and a strongly elliptical shape ($\epsilon>0.1$).
This demonstrates that elliptical groups and clusters grow faster on average than spherical ones.
The Spearman rank correlation coefficient for the ellipticity and instantaneous mass accretion rate, which is about $0.3$ (see Table \ref{tab:correl}), confirms the strong correlation between ellipticity and accretion rate \citep[also recently shown by][]{Lau2021}. This relation of ellipticity and accretion is significant for all masses ranging between $10^{13}$ and $10^{15} M_{\odot}/h$ and suggests that high-accretion flows strongly disturb the shapes of groups and clusters \citep[as discussed in][]{Morinaga2020}.

The right panel of Fig. \ref{fig:dmdt} displays the evolution of the mass accretion rate as a function of mass, considering the three bins of connectivity defined above (little, medium, or highly connected halos). It shows for each mass bin that highly connected clusters tend to grow faster at the present time than other groups and clusters.
The high value of the correlation coefficient for connectivity and the instantaneous mass accretion rate (see Table \ref{tab:correl}) also demonstrates that the accretion rate scales with connectivity. We may assume that more mass is infalling because of the large number of filaments to which the halo is connected. Moreover, in concordance with the theoretical investigations of \cite{Musso2018}, our result suggests that at fixed mass, halos in the nodes (with $\kappa>3$) have higher accretion rates than halos in filaments (with $\kappa<3$).

From these results, we conclude that the anisotropy of the halo environment affects their shapes: the larger the number of filaments to which a group or cluster is connected, the more elongated its shape on average.
For the first time, we highlight the correlation between ellipticity and connectivity and show that this is not only driven by the mass. 
Moreover, our results suggest that this link between anisotropic environment and halo shape can be due to their current growth. Regardless of their mass, connectivity and ellipticity are strongly correlated to the accretion of groups and clusters at the present time. Our results reveal that clusters that are highly connected to the cosmic web tend to grow faster, and these fast accretion flows must disturb their mass distribution, hence increase their ellipticity.

\subsection{Effect of the dynamical state on group anisotropies \label{RES:DS}}

\begin{figure*}
    \centering
    \includegraphics[width=0.85\textwidth]{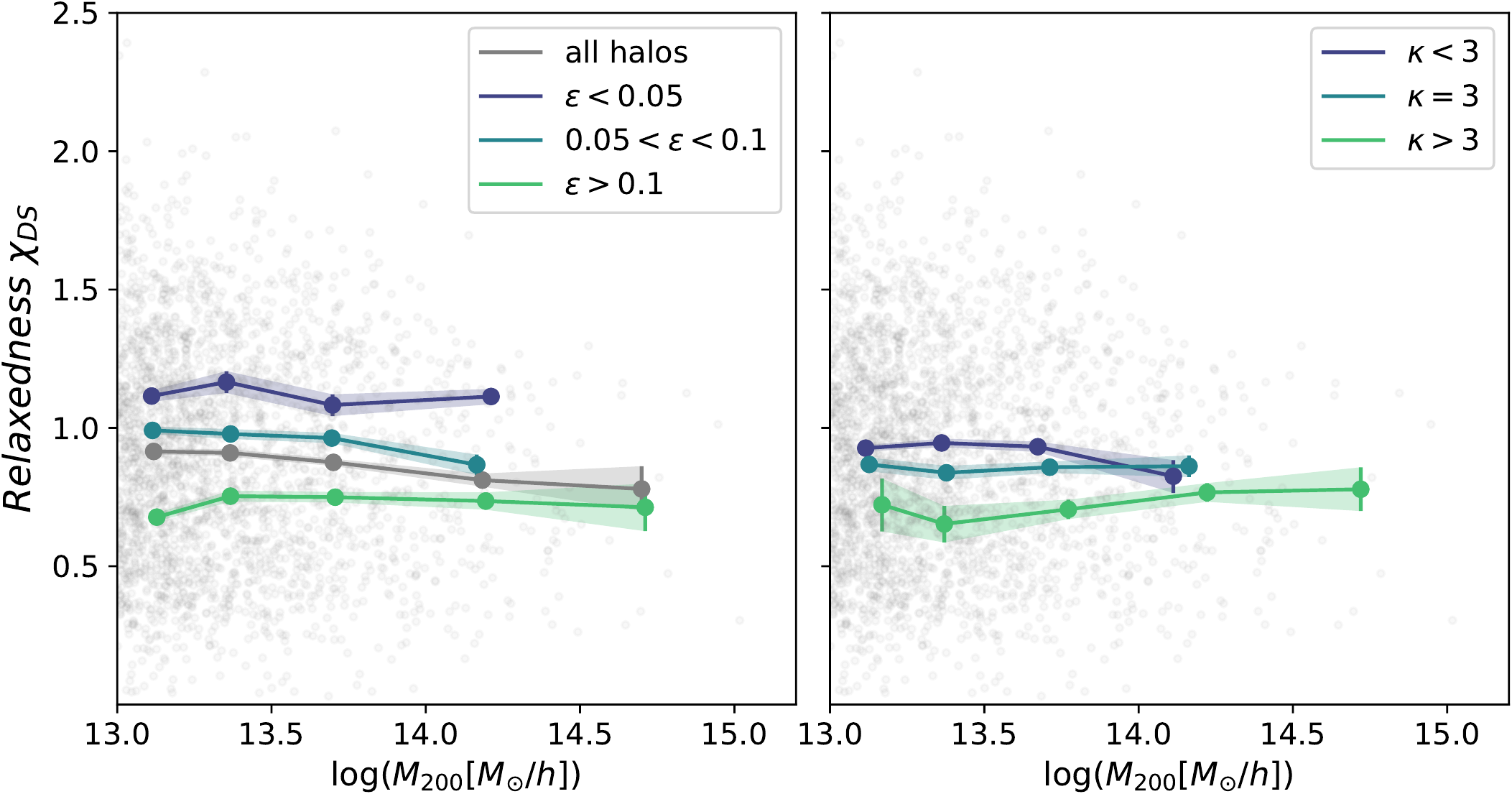}
    \caption{ Mean relaxedness (computed with Eq. \ref{EQ:relax}) as a function of mass. In the right panel, the mean relaxedness is shown for three different ellipticity bins: almost spherical ($\epsilon<0.05$), quite elliptical ($0.05<\epsilon<0.1$), and elliptical ($\epsilon>0.1$). The gray line traces the mean relaxedness for all groups and clusters, regardless of their ellipticity. In the left panel, the mean relaxedness is shown for four different connectivity bins: $\kappa=1$, $\kappa=2$, $\kappa=3$,  and $\kappa>3$. 
    The error bars are the errors on the mean, derived from bootstrap resampling.
    \label{fig:dynamic}}
\end{figure*}

To proceed in probing possible correlations between properties of groups and clusters and their environment, we investigated the effect of the dynamical state at $z=0$.
The dynamical state is quantified by the relaxation level, relaxedness, using the three proxies defined in Sect. \ref{SECT:DS}: the center offset, the virial ratio, and the fraction of substructure.
Figure \ref{fig:dynamic} shows the mean relaxedness $\chi_\text{DS}$ as a function of mass for three different ellipticity bins (right panel) and three different connectivity bins (left panel).
The relation between relaxation and mass is displayed in the left panel of Fig. \ref{fig:dynamic} for all objects (independently of their ellipticity and connectivity). As expected, the level of relaxation decreases with mass, such that massive structures are less relaxed on average than low-mass structures \citep[see also][]{Kuchner2020}. The fact that the halo mass drives the dynamical state is indeed rather well established in theory based on large N-body simulation \citep[see, e.g.,][]{Power2012}. This can be explained by the fact that high-mass halos tend to form later and are still in formation phases \citep{Giocoli2012}. 
In our study, we focus on a correlation in addition to the mass-driven effect.

In addition to this mass-relaxedness relation, we found that the large scatter in relaxedness can be attributed to the different shapes and connectivities that characterize groups and clusters.
The left panel of Fig. \ref{fig:dynamic} shows that elliptical structures are less relaxed than spherical structures in all mass bins. Table \ref{tab:correl} shows that the high value of the correlation coefficient between ellipticity and relaxedness confirms that the shape is a good tracer of the dynamical state of clusters.

The right panel of Fig. \ref{fig:dynamic} displays the relaxedness-mass relation for different connectivity values. The correlation coefficient $\rho_\text{sp} (\kappa,\chi_\text{DS}) = 0.13$ indicates a weak anticorrelation between relaxedness and connectivity. This value tends to increase when low-mass halos are excluded (see Table \ref{tab:correl}).
At fixed mass, the right panel of Fig. \ref{fig:dynamic} shows that weakly connected groups and clusters are on average more relaxed than highly connected groups and clusters in all mass bins. This result argues in favor of a large-scale environmental dependence on the dynamical properties of groups and clusters. 

\begin{figure*}
    \centering
    \includegraphics[width=0.85\textwidth]{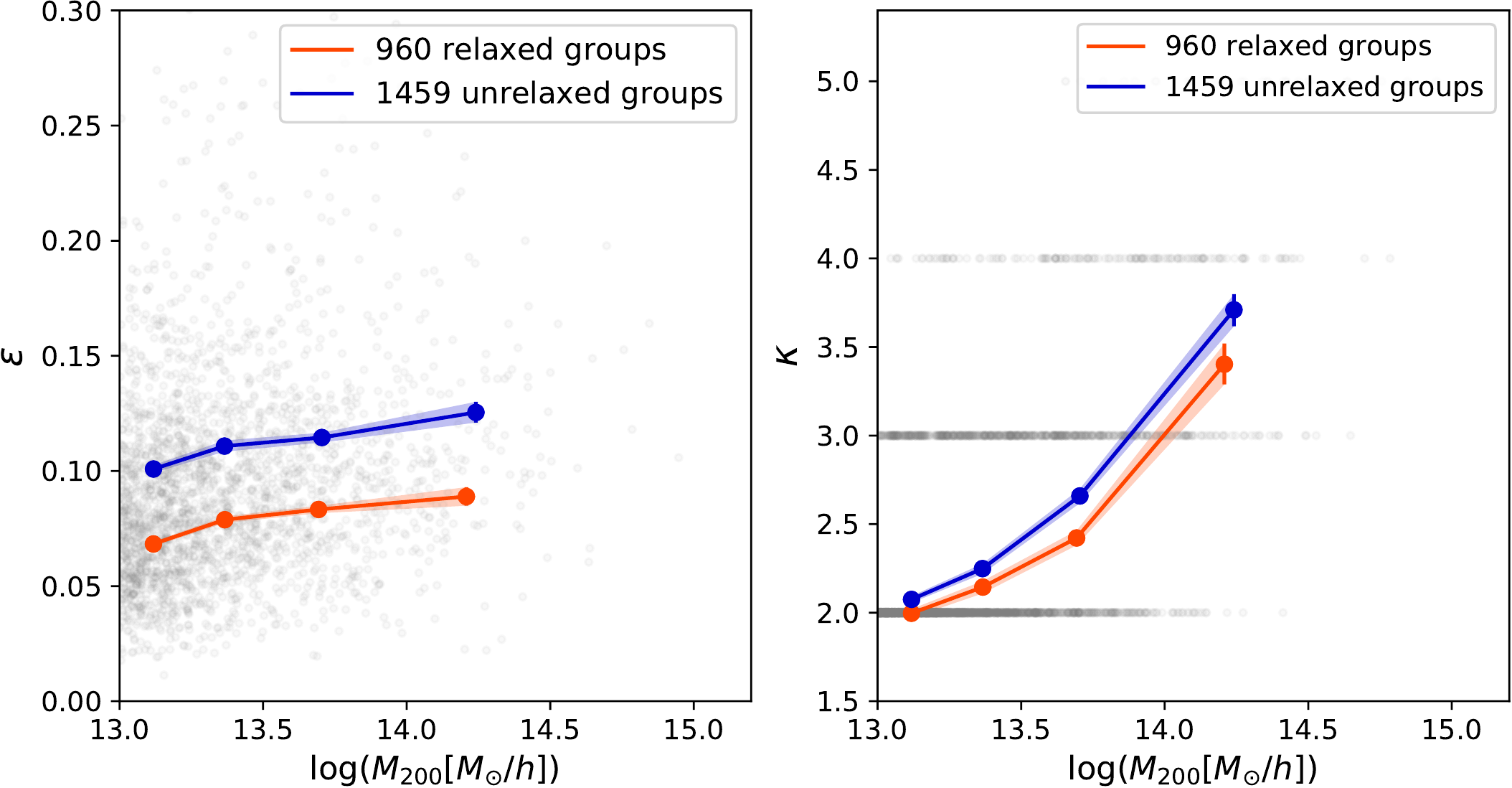}
   \caption{Mean ellipticity (right panel) and mean connectivity (left panel) as a function of mass, separately considering relaxed and unrelaxed groups and clusters (relaxed groups have a relaxedness $\chi_\text{DS} \geq 1$).
   The error bars are the errors on the mean, derived from bootstrap resampling. In gray we show the point distribution of the sample.
   \label{fig:dynamic2}}
\end{figure*}

Another way to explore the link between the dynamical state and anisotropy proxies is to separately consider the ellipticity- and connectivity-mass relations for relaxed and unrelaxed clusters, as presented in the panels of Fig. \ref{fig:dynamic2}. As discussed in Sect \ref{SECT:DS}, groups and clusters are considered as relaxed if the relaxnedness condition $\chi_\text{DS} \geq 1$ is fulfilled. 
The left panel of Fig. \ref{fig:dynamic2} shows that relaxed structures are more spherical on average than unrelaxed structurse in each mass bin.
This agrees with the result of \cite{Suto2016}, who found that halos populated by massive substructures are systematically less spherical than  single halos. 
The right panel of Fig. \ref{fig:dynamic2} highlights the fact that unrelaxed groups and clusters are also more strongly connected to the cosmic web (i.e., with a higher value of connectivity) than relaxed groups and clusters. This result is significant for each mass bin. It consequently shows that regardless of the mass, the connectivity and ellipticity tend to trace different dynamical states of groups and clusters.

Finally, we conclude that in addition to the mass effect on halo relaxation, internal and external indicators of anisotropy trace the dynamical state. Regardless of the mass, dynamically unrelaxed groups and clusters are more strongly connected to the cosmic web on average and are more elliptical than the relaxed groups and clusters. This second result, in addition to our previous finding on the effect of cosmic filament accretion on shapes, suggests that highly connected systems are strongly accreting matter, which in turn perturbs their matter distribution, and thus affects both their dynamical state and shape.

\subsection{Effect of mass growth history on anisotropy \label{Sect_res_final}}

We found that both ellipticity and connectivity trace the dynamical state and the recent growth of groups and clusters.
We now explore further whether these relations of anisotropy proxies and halo dynamics are the result of different growth histories.
Previous studies have postulated that dynamical states of halos can be attributed to their mass-assembly history (MAH) \citep[see, e.g., ][]{Power2012, Mostoghiu2019}.
Therefore we expect that the MAH also affects the anisotropy of the density field at $z=0$. 
We considered two quantities to describe the MAH of groups and clusters: the continuous mass accretion rate $\Gamma_{200\text{m}}(z=0.5)$, and the formation redshift $z_\text{form}$, as defined in Sect. \ref{SECT:MAR}.

\begin{figure*}
    \centering
    \includegraphics[width=0.85\textwidth]{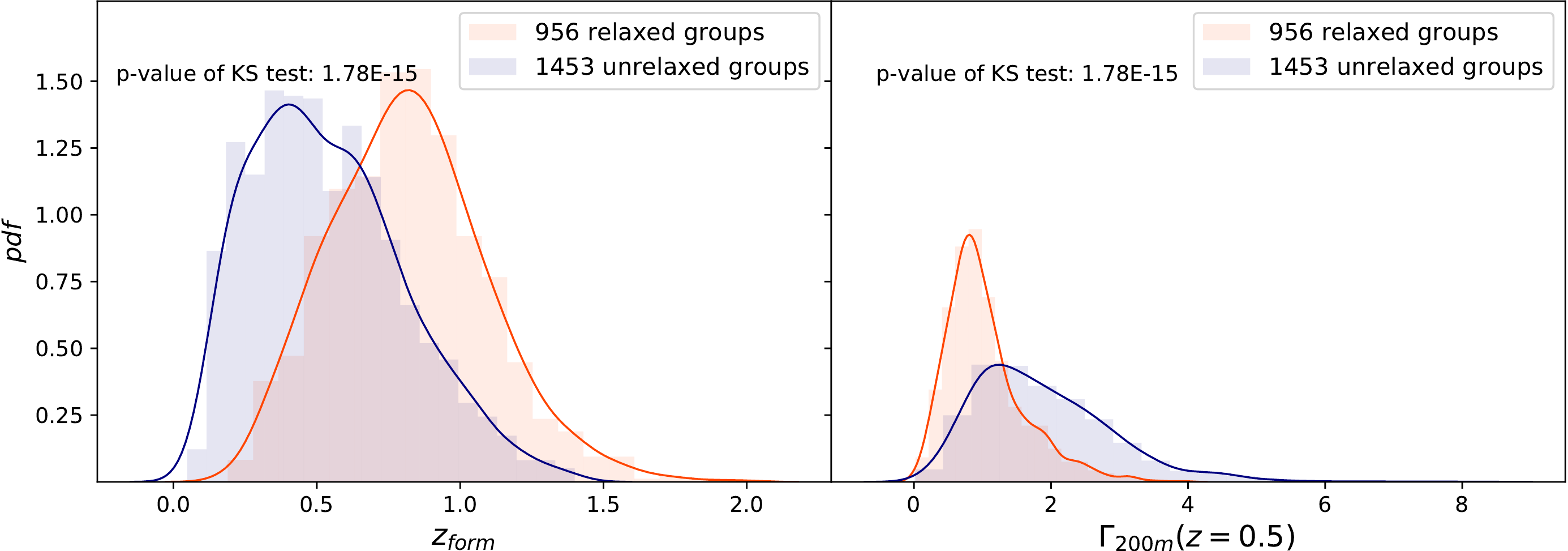}
    \caption{Probability distribution function of the continuous accretion rate $\Gamma_{200\text{m}}$ and the formation redshift $z_\text{form}$ for relaxed (red) and unrelaxed (blue) groups and clusters. The low \textit{p}-value of KS test (at the top of panels) confirms that relaxed and unrelaxed objects are two distinct populations in terms  of their mass-assembly history. \label{fig:histogramm_zf_MAR}}
\end{figure*}

In Fig. \ref{fig:histogramm_zf_MAR} we plot the probability distribution function of the mass accretion rate $\Gamma_{200m}(z=0.5)$ (left panel) and the formation redshift $z_\text{form}$ (right panel) for relaxed and unrelaxed groups and clusters separately. We clearly observe that relaxed and unrelaxed structures are two different populations with different distributions of formation time and accretion rate.
In agreement with \cite{Power2012}, we find that young objects are more likely to be dynamically unrelaxed than their older counterparts (in the right panel). Moreover, the left panel illustrates the fact that relaxed structures are on average in a slower accretion phase than unrelaxed structures.

\begin{figure*}
    \centering
    \includegraphics[width=0.75\textwidth]{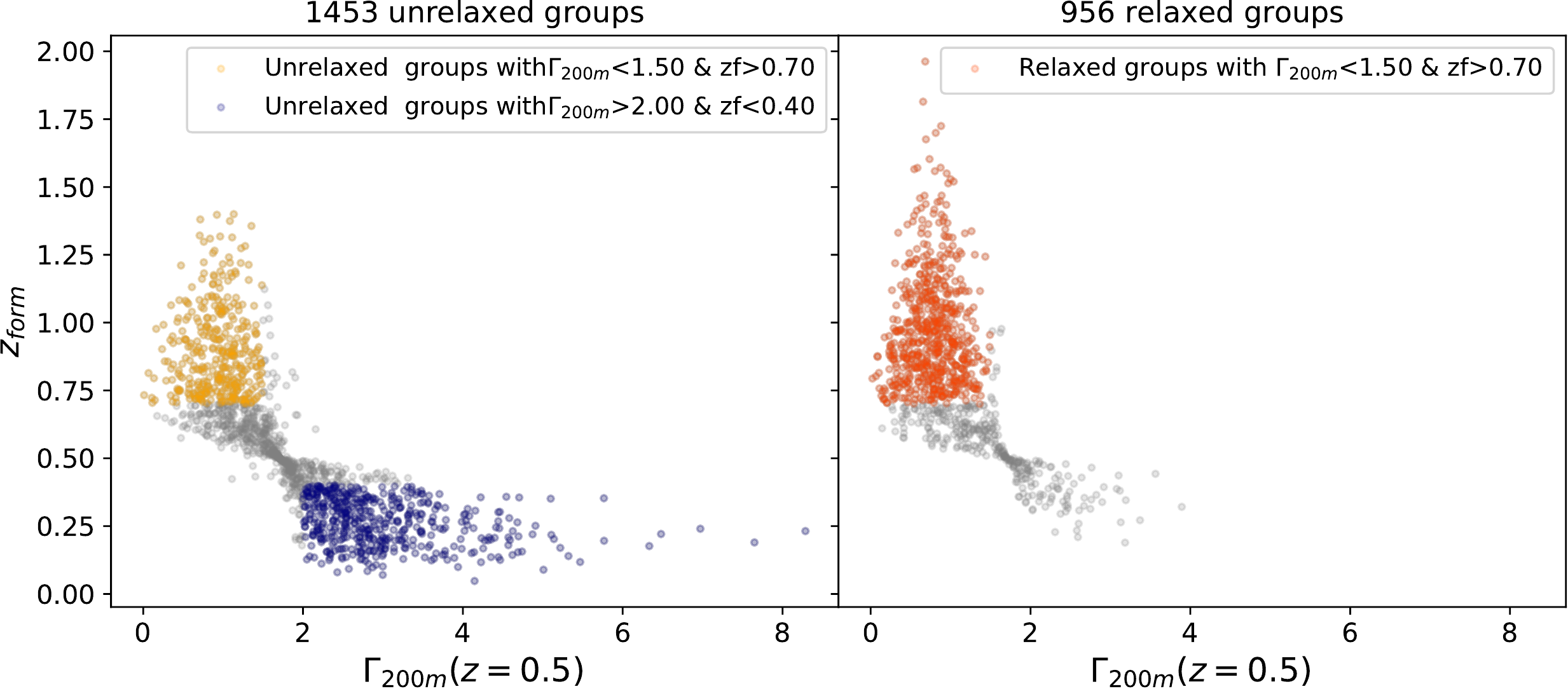}
    \caption{ Distribution of groups and clusters in formation redshift and mass accretion rate ($z_\text{form}$,$\Gamma_{200\text{m}}$) space for relaxed and unrelaxed populations separately. We highlight three main subsamples: 620 early-formed low-accreting and relaxed groups and clusters in red, 360 early-formed low-accreting and unrelaxed in yellow, and 482 late-formed fast-accreting and unrelaxed groups and clusters in blue. In gray we show the overall point distribution of the full sample. \label{fig:hist_final} }
\end{figure*}

In order to select groups and clusters according to their MAH, we display in Fig. \ref{fig:hist_final} the distribution of formation redshift as a function of the continuous mass accretion rate for unrelaxed (left panel) and relaxed structures (right panel).
Regardless of the dynamical state, we observe that $z_\text{form}$ and $\Gamma_{200\text{m}}$ are strongly correlated: Recently formed groups and clusters are in a faster accretion phase than early formed groups and structures. This relation of the formation time and accretion rate is caused by their definition; both are computed from the halo mass growth $M_{\mathrm{200m}}(t)$ as defined in Sect \ref{SECT:MAR}. These two MAH proxies are in principle related through the functional form proposed in \cite{Wechsler2002} such that $M_{200}(z)= M_{200}(z=0) \exp(-\alpha z)$ with $\alpha$  parametrizing the mass accretion rate of the halo.
Our two MAH proxies, formation time and accretion rate, are assumed to account for only $\sim 70 \%$ of the scatter in mass accretion histories because of the stochastic effect of major mergers \cite[as discussed by][]{Wong2012,Chen2020}. This explains the degeneracy in the formation time - accretion rate relation that is plotted in Fig. \ref{fig:hist_final}.

\begin{figure*}
    \centering
    \includegraphics[width=0.85\textwidth]{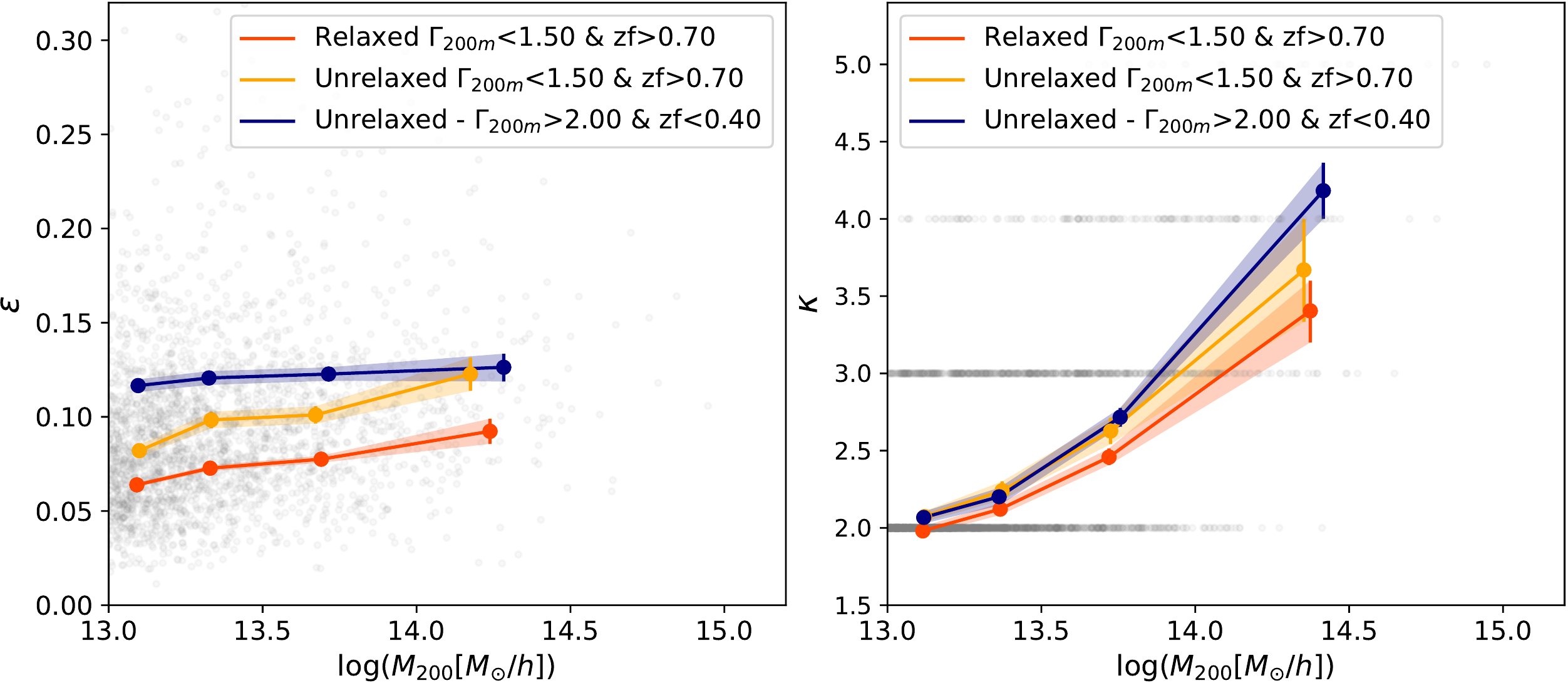} 
    \caption{Mean ellipticity (right panel) and mean connectivity (left panel) as a function of the mass, separately considering three subsamples of groups and clusters with different mass assembly histories and dynamical status: 620 early-formed low-accreting and relaxed in red, 360 early-formed low-accreting and unrelaxed in yellow, and 482 late-formed fast-accreting and unrelaxed in blue. In gray we show the point distribution of the full sample. \label{fig:FINAL_fig} }
\end{figure*}

Fig. \ref{fig:FINAL_fig} shows that relaxed and unrelaxed groups and clusters have different ($z_{\mathrm{form}}$,$\Gamma_{\mathrm{200m}}$) distributions, with typically three main \textup{}evolutionary-dynamical populations. We highlight these populations by considering cuts in the tails of mass accretion and formation time distributions, as illustrated in Fig. \ref{fig:FINAL_fig}. Relaxed objects are formed long time ago and they are slowly accreting matter in their recent history. We also show unrelaxed halos that are also early-formed and low-accreting. The third population is constituted of late-formed unrelaxed halos that are in fast accretion phase. A small number of relaxed structures are formed late and accrete fast, but because they are rare, we did not consider them here.

Focusing on these three \textit{\textup{evolutionary-dynamical populations}} of groups and clusters, we explore their anisotropy proxies in Fig. \ref{fig:FINAL_fig}. These populations are characterized by different anisotropies, that is, ellipticity and connectivity of groups and clusters, regardless of their mass. 
First, old and relaxed halos are rather spherical and weakly connected to their environment. These groups might have formed early enough to relax and thus lose memory of the directions of the accretion and mergers they experienced.
Second, the population of the early-formed but still unrelaxed groups and clusters at the present time is characterized by intermediate connectivity and ellipticity values. This old unrelaxed sample of groups might be still in a transition\textup{} phase, that is, the halos did not have enough time to relax. They could be post-merger, that is, the result of merging systems that have not completely achieved their merger at the present time.
Finally, the late-formed unrelaxed sample of clusters has significantly more anisotropic shapes, with a higher ellipticity value, and is highly connected to the large-scale cosmic web. These groups and clusters must typically be in formation phase and be strongly affected by the infalling material through connected filaments.
These young unrelaxed structures might be merging or recent post-merging systems.
Fast mass growth is dominated by major mergers (as detailed in \citep{Valles2020}).
Moreover, as suggested by \cite{Darragh2019}, groups with the highest connectivities have encountered a major merger more recently than structures with lower connectivity.
In agreement with these interpretations, previous studies have shown that unrelaxed halos tend to be the result of one or more recent significant mergers \citep[see, e.g.,][]{Hetznecker2006,DOnghia2007,Power2012,Klypin2016}.
We conclude that regardless of the mass, systems still in formation phase (recently formed and in a high-accretion phase) are significantly more strongly connected to the cosmic web and have more disturbed and elliptical shapes on average than early-formed relaxed systems.

\section{Discussion \label{DISCUSION}}

Halo growth is commonly explored through the evolution of the halo density profile.
Recently, \cite{Mostoghiu2019} have studied the density profiles of simulated clusters as a function of their mass-assembly history. They found that early-formed relaxed clusters have entered a slow-accretion phase, therefore their mass growth mostly occurs in their outskirts. In contrast, the authors showed that late-formed unrelaxed clusters are in their fast-accretion phase and their central regions are still growing.
These results complement our analysis of the overall picture of cluster evolution well and show that young unrelaxed clusters are still growing in their center and thus appear nonspherical in our study, whereas old relaxed clusters are more spherical because their centers no longer grow.

In addition, it is important to recall that both the evolutionary and the dynamical states of groups and clusters are most of all driven by their masses. The fraction of objects that are recently formed and/or are unrelaxed increases drastically with mass. This is the direct consequence of the hierarchical structure formation scenario in which most of the massive systems are still in formation phase because of their high accretion and recent merger activities \citep[as detailed in][e.g.]{Power2012, Giocoli2012}.
Therefore this hierarchical mass assembly model explains the main effect of the mass in shaping the density field of group and cluster environments. We found that the ellipticity and connectivity are higher for massive halos because they are mostly recently formed systems \citep[also discussed in][for the ellipticity]{Despali2014}.
Thus, our work tends to indicate that the mass assembly bias is indeed directly related to the intrinsic anisotropic nature of the density field imposed by the cosmic web. In agreement with \cite{Musso2018}, our results suggest that older halos with lower mass preferentially lie inside filaments or low-connected structures, while more massive and younger halos are located in the highly connected nodes. In addition to the trend governed by the mass assembly, we showed the additional contribution of the halo accretion history to the halo anisotropies.

Finally, it is important to keep in mind that our results and discussions of the shape and connectivity of groups and clusters are theoretical predictions based on cosmological simulations. 
In observations, the measurement of group and cluster ellipticities remains uncertain given observational limitations (projection effects, resolution, signal-to-noise ratio, etc.). It is hence difficult to compare measured morphological indicators from different observables (X-ray, Sunyaev-Zeldovich effect, galaxy distribution, or lensing) with each other and with theoretical predictions.  Gravitational lensing provides a unique technique for constraining the ellipticity of the total matter distribution that is directly comparable with theoretical predictions.
Measuring the mean projected halo ellipticity from two-dimensional lensing shear maps, \cite{Oguri2010} reported the measurement of the ellipticity of $\sim 20$ clusters and found an excellent agreement with theoretical predictions \citep[see also][and citations within]{Evans2009, Clampitt2016B}. Recently, \cite{Gonzalez2021} measured the shape of the projected surface mass density distribution of redMaPPer galaxy clusters \citep{Redmapper} using a weak-lensing stacking technique, but did not find the expected mass-ellipticity relation from simulations. The authors discussed the possibility that the effects introduced by the contribution of the surrounding halos on the line of sight might induce an underestimation of the ellipticity, which mainly affects the high-mass halos.
To the best of our knowledge, other cluster ellipticity measurements based on observations, such as using galaxy members \citep{Shin2018}, do not succeed in finding significant observational evidence that more massive clusters are more elliptical at the present time.

\par\bigskip
Regarding the relation between the dynamical state and the shape of clusters, \cite{Bartalucci2019}  used the sparsity as a morphology proxy to recently highlight evidence that the shape of the total mass profile in massive clusters is governed by their dynamical state. In addition, \cite{Yuan2020} probed the dynamical state for 964 galaxy clusters from Chandra X-ray images and reported strong correlations of the X-ray morphology index and optical dynamical relaxation factor \citep[see also][]{Rossetti2016}. Nevertheless, it is important to note that the observational results based on both X-ray and Sunyaev-Zeldovich effect (SZ) only probe the intracluster medium of groups and clusters and mainly focus on the cluster core ($<R_{500}$). 
Moreover, given the collisional nature of the gaseous component of clusters, the gas mass distribution (and hence its X-ray surface brightness and the Compton $y$-parameter) is significantly more circular than the stellar and dark matter counterparts \citep[as predicted by][using hydrodynamical simulations]{Okabe2018, Harvey2021}. Moreover, 
simulations show that the shapes of different cluster-component distributions (stars, gas, and dark matter) qualitatively follow similar trends with halo mass, radius, and redshift \citep{Velliscig2015, Okabe2018}.
 \cite{Donahue2016} have confirmed this strong correlation of the morphological properties measured from X-ray (for the gas) and lensing maps (for the total mass) for $\sim 20$ CLASH galaxy clusters. 
In order to reach consensus about the classification of clusters in simulations and observations, \cite{DeLuca2020} explored the correlations of morphology indicators and dynamical state proxies extracted from both simulated clusters and their X-ray and Compton $y$-parameter mock images. The authors concluded that the offset parameters based on the centroids of X-ray and SZ maps are more sensitive to the dynamical state than to peak position offsets.

\begin{figure}
   \centering
   \includegraphics[width=0.45\textwidth]{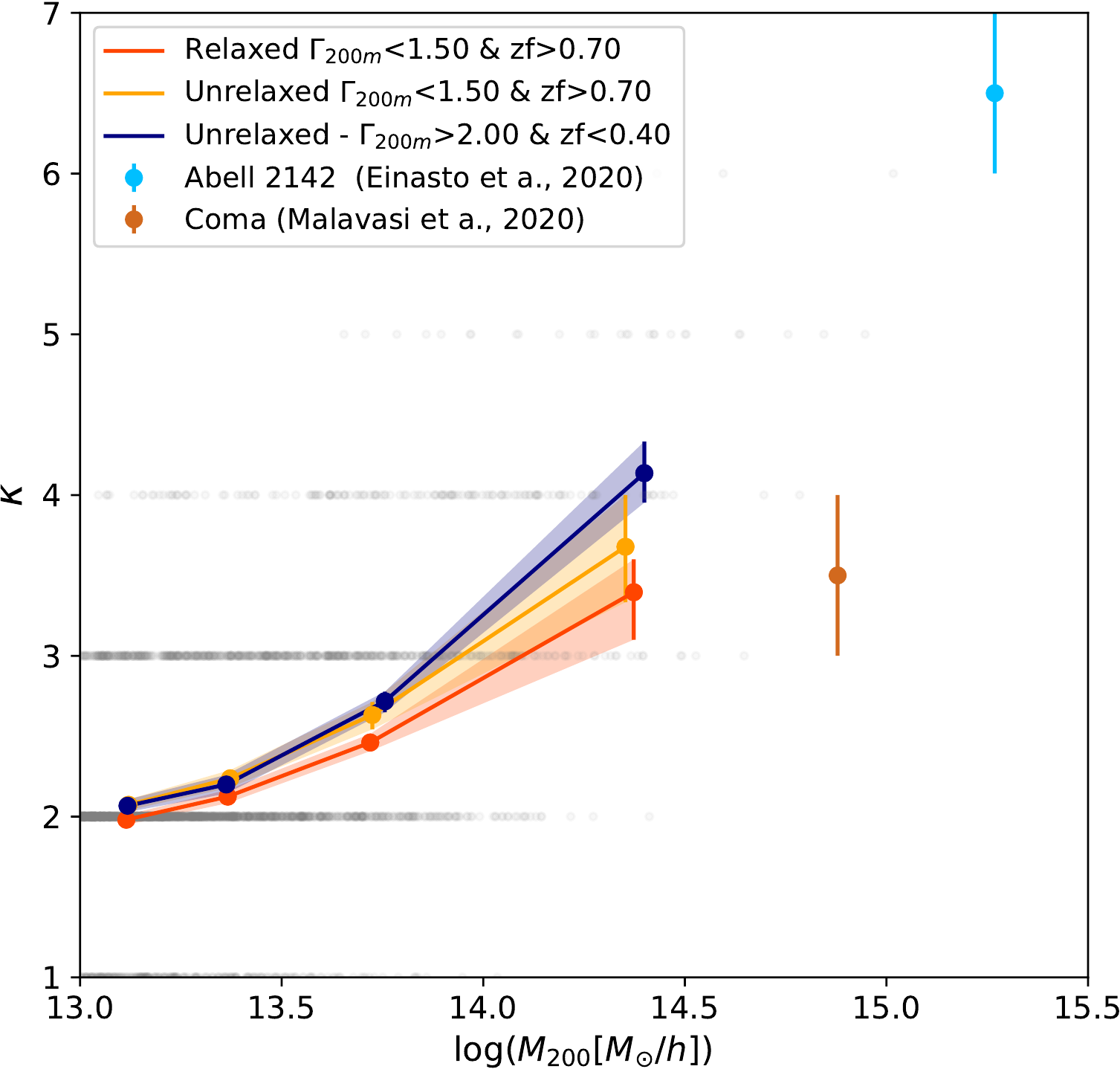}
   \caption{Mean connectivity as a function of the two main populations in MAH and dynamical state, as discussed in Sect. \ref{Sect_res_final}. The two observational measurements of A2142 and the Coma cluster are shown in cyan and brown, respectively. \label{fig:OBS} }
\end{figure}

By exploring the connection of clusters and their large-scale environment in observations, the theoretical expectation that more massive halos are more strongly connected to the cosmic web has been confirmed from large galaxy surveys \citep[see, e.g., ][]{Darragh2019, Kraljic2020}.
Concerning the relation of halo environments and their dynamical state, no observational studies have statistically explored the relation of the cluster dynamical state and their connectivity so far. Only two peculiar galaxy clusters for which the connectivity and dynamical state have been individually estimated from observations can be used to illustrate our findings. 
First, \cite{Einasto2015,Einasto2018,Einasto2018B,Einasto2020} showed that the A2142 cluster is dynamically unrelaxed with many massive substructures that are highly connected to its large-scale environment and with an elongated shape \citep[see also][for the perturbed dynamics of A2142]{Rossetti2013}.
The mass of the A2142 cluster is estimated to be $M_{200} = 1.3 \times 10^{15} M_{\odot}$ \citep{Munari2014} and is connected to approximately six or seven filaments \citep{Einasto2020}.
Second, the well-studied Coma cluster \citep[see][for a review]{Biviano1998} is weakly connected to other structures in the field \citep{Mahajan2018} with a connectivity estimated between 3 and 4, as measured by \cite{Malavasi2020}. The Coma cluster seems to be in a low-accretion phase at the present time, with a mass growth of around $1/5$ since $z\sim0.2 - 0.3$ \citep{Adami2005}.
However, the dynamical state of the Coma cluster is not completely established. While initially, it was classified as a relaxed system \citep{Kent1982} with a very high concentration parameter suggesting a compact and old object \citep{Lokas2003}, other observational evidence has argued in favor of a dynamically unrelaxed state: two BCGs \citep{Fitchett1987}, no cool core \citep{Neumann2003}, and its radio halo emission \citep{Thierbach2003}. In addition, \cite{Gavazzi2009} estimated that the mass of the Coma cluster is equal to $M_{200} = 5.3 \times 10^{14} M_{\odot}$.
The two observational measurements of the connectivity $\kappa$ and mass of A2142 and the Coma cluster are plotted in Fig. \ref{fig:OBS}.
Even though we did not statistically probe this very high mass range, the Coma cluster shows the connectivity-mass relation of an early-formed halo well (both relaxed and unrelaxed), while the A2142 cluster appears to follow the relation of late-formed unrelaxed halos that are still in the formation phase.

\section{Conclusion}

It is crucial to study the anisotropic matter distribution inside and around density peaks to understand the formation and growth of massive structures.
We have studied two proxies of the anisotropic matter distribution around halos: the shape (through the ellipticity), and the filamentary structure around them (through the connectivity). We probed these anisotropies for a large sample of galaxy groups and clusters with masses $>1 \times 10^{13} M_{\odot}/h$ in the IllustrisTNG hydrodynamical simulation at $z=0$ \citep{ILLUSTRIS_TNG}.
The halo shape was approximated by a triaxial ellipsoid that encloses $M_{200}$ and was quantified by the ellipticity parameter, following the definition of \cite{Suto2016}.

In the outskirts of groups and clusters, the anisotropic matter distribution was estimated by quantifying the number of cosmic filaments connected to halos, that is, the local connectivity. The filamentary structure detection algorithm T-ReX \citep{Bonnaire2020} was used to trace the cosmic web skeleton from the galaxy distribution in the simulation box. We then computed the connectivity by counting the number of filaments that are connected to a given halo within a sphere of $1.5 \times R_{200}$.

We confirm that the halo mass is strongly related to the anisotropies in the density field. Namely, massive groups and clusters are on average significantly more elliptical \citep{Allgood2006} and more strongly connected to the cosmic web \citep{Codis2018} than low-mass groups and clusters. 
In additiont to the mass-driven effect, we explored additional correlations between connectivity and ellipticity, and the dynamical state and assembly history of groups and clusters of galaxies.
Our major findings are summarized below.\begin{itemize}
\item[(1)] In addition to the mass-driven effect, ellipticity and connectivity are correlated, such that highly connected groups and clusters tend to be more elliptical than weakly connected groups and clusters.

\smallskip

\item[(2)]  The growth rate of groups and clusters is strongly correlated with the ellipticity and connectivity: Fast-accreting objects are on average more elliptical and more strongly connected to the cosmic web than low-accreting objects. This indicates that strongly connected objects are more elliptical because they accrete matter more rapidly, which in turn must disturb the associated matter distribution. Our results also indicate that the accretion rate scales with the connectivity such that at fixed mass, halos in nodes have a higher accretion rate than halos in filaments \citep[in agreement with the theoretical investigations of][]{Musso2018}.

\smallskip

\item[(3)] Both anisotropy proxies trace the dynamical state of halos: Quasi-spherical and weakly connected groups and clusters are strongly relaxed. In contrast, groups and clusters that are dynamically unrelaxed are on average significantly more elliptical and more connected. This trend holds for all mass bins. 
The large scatter in the well-known relaxedness-mass relation can indeed be attributed to the different anisotropic level in the density field around groups and clusters.

\smallskip

\item[(4)] We postulate that the relation between dynamical state and anisotropy proxies can be attributed to the different mass-assembly histories of halos. Relaxed and unrelaxed groups and clusters are two distinct populations. The dynamically unrelaxed gropus and clusters formed more recently on avearage and are in a faster accreting phase than relaxed halos. Our comprehensive analysis of the halo accretion history and dynamical state showed that there are mainly two distinct populations: the early-formed slowly accreting relaxed halo sample, and the late-formed fast-accreting unrelaxed halos. 
These two types of \textup{}evolutionary-dynamical states of halos present different anisotropies in their density field.
Old relaxed groups and clusters acrrete matter slowly, are almost spherical, and have a weak connection to their environment. They should have lost their connection with their preferential directions of accretion and merging.
In contrast, young fast-accreting and unrelaxed groups and clusters are much more elliptical and strongly connected to the overall cosmic web. This population must be in formation phase and strongly affected by the environment that feeds them abundantly.
The two populations defined according to their evolutionary-dnymical state are significantly different in all mass bins, even if the fraction of recently formed objects increases with mass, in agreement with the hierarchical structure formation scenario. 
\end{itemize}

With this first comprehensive analysis, we open new ways for relating the growth, dynamical state, and the shape of groups and clusters with their connection to the large-scale structure.
Our results show that different recent accretion histories lead to different anisotropies in the matter distribution (with both distinct ellipticity and connectivity) as a second driving effect after the well-known mass effect.
Our analyses show that the correlation between connectivity and halo properties is more significant for higher-mass clusters, given that low-mass groups are mainly located inside cosmic filaments \citep{Cautun2014}. We conclude that to significantly detect correlations of halo connectivity and halo properties, we need to focus on clusters with masses $M_{200}>10^{13.5} M_{\odot}/h$. By probing a larger sample of simulated clusters ($M_{200}>10^{14} M_{\odot}/h$), which are ideal structures for probing a wider range of connectivities, we intend to further explore the cosmic time evolution in the relations of the large-scale environment and cluster properties in a future work.

From the observational point of view, the A2142 and Coma clusters both provide illustrations of two distinct cluster connectivities that might be the result of different mass-assembly histories.
The new generation of large galaxy and cluster surveys, such as \textit{Euclid} and \textit{Dark Energy Survey}, will enable us to probe many massive clusters and to investigate the effect of large-scale structure connections on the dynamical properties of the halos. 
In addition, upcoming accurate multiwavelength observations of galaxy clusters such as the Cluster HEritage project \citep{Heritage2020} will allow us to probe the dynamical properties of clusters and their matter distributions from their central regions to their outskirts. They wil thus be used to test our theoretical predictions.
\begin{acknowledgements}

The authors acknowledge the useful comments of the anonymous referee.
This research has been supported by the funding for the ByoPiC project from the European Research Council (ERC) under the European Union’s Horizon 2020 research and innovation program grant agreement ERC-2015-AdG 695561 (ByoPiC, https://byopic.eu). The authors thank the very useful comments and discussions with all the members of the ByoPiC team. The authors acknowledge the use of the IllustrisTNG publicly available data.

\end{acknowledgements}

\bibliographystyle{aa}

\begin{thebibliography}{103}
\expandafter\ifx\csname natexlab\endcsname\relax\def\natexlab#1{#1}\fi

\bibitem[{{Adami} {et~al.}(2005){Adami}, {Biviano}, {Durret}, \&
  {Mazure}}]{Adami2005}
{Adami}, C., {Biviano}, A., {Durret}, F., \& {Mazure}, A. 2005, \aap, 443, 17

\bibitem[{{Allgood} {et~al.}(2006){Allgood}, {Flores}, {Primack}, {Kravtsov},
  {Wechsler}, {Faltenbacher}, \& {Bullock}}]{Allgood2006}
{Allgood}, B., {Flores}, R.~A., {Primack}, J.~R., {et~al.} 2006, \mnras, 367,
  1781

\bibitem[{{Arag{\'o}n-Calvo} {et~al.}(2010){Arag{\'o}n-Calvo}, {van de
  Weygaert}, \& {Jones}}]{Aragon2010}
{Arag{\'o}n-Calvo}, M.~A., {van de Weygaert}, R., \& {Jones}, B. J.~T. 2010,
  \mnras, 408, 2163

\bibitem[{{Bardeen} {et~al.}(1986){Bardeen}, {Bond}, {Kaiser}, \&
  {Szalay}}]{Bardeen1986}
{Bardeen}, J.~M., {Bond}, J.~R., {Kaiser}, N., \& {Szalay}, A.~S. 1986, \apj,
  304, 15

\bibitem[{{Bartalucci} {et~al.}(2019){Bartalucci}, {Arnaud}, {Pratt},
  {D{\'e}mocl{\`e}s}, \& {Lovisari}}]{Bartalucci2019}
{Bartalucci}, I., {Arnaud}, M., {Pratt}, G.~W., {D{\'e}mocl{\`e}s}, J., \&
  {Lovisari}, L. 2019, \aap, 628, A86

\bibitem[{{Biviano}(1998)}]{Biviano1998}
{Biviano}, A. 1998, in Untangling Coma Berenices: A New Vision of an Old
  Cluster, ed. A.~{Mazure}, F.~{Casoli}, F.~{Durret}, \& D.~{Gerbal}, 1

\bibitem[{{Bonamigo} {et~al.}(2015){Bonamigo}, {Despali}, {Limousin}, {Angulo},
  {Giocoli}, \& {Soucail}}]{Bonamigo2015}
{Bonamigo}, M., {Despali}, G., {Limousin}, M., {et~al.} 2015, \mnras, 449, 3171

\bibitem[{{Bonnaire} {et~al.}(2020){Bonnaire}, {Aghanim}, {Decelle}, \&
  {Douspis}}]{Bonnaire2020}
{Bonnaire}, T., {Aghanim}, N., {Decelle}, A., \& {Douspis}, M. 2020, \aap, 637,
  A18

\bibitem[{{Cadiou} {et~al.}(2020){Cadiou}, {Pichon}, {Codis}, {Musso},
  {Pogosyan}, {Dubois}, {Cardoso}, \& {Prunet}}]{Cadiou2020}
{Cadiou}, C., {Pichon}, C., {Codis}, S., {et~al.} 2020, \mnras, 496, 4787

\bibitem[{{Cautun} {et~al.}(2014){Cautun}, {van de Weygaert}, {Jones}, \&
  {Frenk}}]{Cautun2014}
{Cautun}, M., {van de Weygaert}, R., {Jones}, B. J.~T., \& {Frenk}, C.~S. 2014,
  \mnras, 441, 2923

\bibitem[{{Chen} {et~al.}(2019){Chen}, {Avestruz}, {Kravtsov}, {Lau}, \&
  {Nagai}}]{Chen2019}
{Chen}, H., {Avestruz}, C., {Kravtsov}, A.~V., {Lau}, E.~T., \& {Nagai}, D.
  2019, arXiv e-prints, arXiv:1903.08662

\bibitem[{{Chen} {et~al.}(2020){Chen}, {Mo}, {Li}, {Wang}, {Yang}, {Zhang}, \&
  {Wang}}]{Chen2020}
{Chen}, Y., {Mo}, H.~J., {Li}, C., {et~al.} 2020, \apj, 899, 81

\bibitem[{{Clampitt} \& {Jain}(2016)}]{Clampitt2016B}
{Clampitt}, J. \& {Jain}, B. 2016, \mnras, 457, 4135

\bibitem[{{Codis} {et~al.}(2018){Codis}, {Pogosyan}, \& {Pichon}}]{Codis2018}
{Codis}, S., {Pogosyan}, D., \& {Pichon}, C. 2018, \mnras, 479, 973

\bibitem[{{Cole} \& {Lacey}(1996)}]{Cole1996}
{Cole}, S. \& {Lacey}, C. 1996, \mnras, 281, 716

\bibitem[{{Contigiani} {et~al.}(2020){Contigiani}, {Bah{\'e}}, \&
  {Hoekstra}}]{Contigiani2020}
{Contigiani}, O., {Bah{\'e}}, Y.~M., \& {Hoekstra}, H. 2020, arXiv e-prints,
  arXiv:2012.01336

\bibitem[{{Cui} {et~al.}(2017){Cui}, {Power}, {Borgani}, {Knebe}, {Lewis},
  {Murante}, \& {Poole}}]{Cui2017}
{Cui}, W., {Power}, C., {Borgani}, S., {et~al.} 2017, \mnras, 464, 2502

\bibitem[{{Darragh-Ford} {et~al.}(2019){Darragh-Ford}, {Laigle}, {Gozaliasl},
  {Pichon}, {Devriendt}, {Slyz}, {Arnouts}, {Dubois}, {Finoguenov},
  {Griffiths}, {Kraljic}, {Pan}, {Peirani}, \& {Sarron}}]{Darragh2019}
{Darragh-Ford}, E., {Laigle}, C., {Gozaliasl}, G., {et~al.} 2019, arXiv
  e-prints, arXiv:1904.11859

\bibitem[{{Dav{\'e}} {et~al.}(2001){Dav{\'e}}, {Spergel}, {Steinhardt}, \&
  {Wandelt}}]{Dave2001}
{Dav{\'e}}, R., {Spergel}, D.~N., {Steinhardt}, P.~J., \& {Wandelt}, B.~D.
  2001, \apj, 547, 574

\bibitem[{{Davis} {et~al.}(1985){Davis}, {Efstathiou}, {Frenk}, \&
  {White}}]{Davis1985}
{Davis}, M., {Efstathiou}, G., {Frenk}, C.~S., \& {White}, S.~D.~M. 1985, \apj,
  292, 371

\bibitem[{{De Luca} {et~al.}(2020){De Luca}, {De Petris}, {Yepes}, {Cui},
  {Knebe}, \& {Rasia}}]{DeLuca2020}
{De Luca}, F., {De Petris}, M., {Yepes}, G., {et~al.} 2020, arXiv e-prints,
  arXiv:2011.09002

\bibitem[{{Despali} {et~al.}(2017){Despali}, {Giocoli}, {Bonamigo}, {Limousin},
  \& {Tormen}}]{Despali2017}
{Despali}, G., {Giocoli}, C., {Bonamigo}, M., {Limousin}, M., \& {Tormen}, G.
  2017, \mnras, 466, 181

\bibitem[{{Despali} {et~al.}(2014){Despali}, {Giocoli}, \&
  {Tormen}}]{Despali2014}
{Despali}, G., {Giocoli}, C., \& {Tormen}, G. 2014, \mnras, 443, 3208

\bibitem[{{Diemer} \& {Kravtsov}(2014)}]{Diemer2014}
{Diemer}, B. \& {Kravtsov}, A.~V. 2014, \apj, 789, 1

\bibitem[{{Diemer} {et~al.}(2013){Diemer}, {More}, \& {Kravtsov}}]{Diemer2013b}
{Diemer}, B., {More}, S., \& {Kravtsov}, A.~V. 2013, \apj, 766, 25

\bibitem[{{Donahue} {et~al.}(2016){Donahue}, {Ettori}, {Rasia}, {Sayers},
  {Zitrin}, {Meneghetti}, {Voit}, {Golwala}, {Czakon}, {Yepes}, {Baldi},
  {Koekemoer}, \& {Postman}}]{Donahue2016}
{Donahue}, M., {Ettori}, S., {Rasia}, E., {et~al.} 2016, \apj, 819, 36

\bibitem[{{D'Onghia} \& {Navarro}(2007)}]{DOnghia2007}
{D'Onghia}, E. \& {Navarro}, J.~F. 2007, \mnras, 380, L58

\bibitem[{{Doroshkevich}(1970)}]{Doroshkevich1970}
{Doroshkevich}, A.~G. 1970, Astrofizika, 6, 581

\bibitem[{{Einasto} {et~al.}(2018{\natexlab{a}}){Einasto}, {Deshev}, {Lietzen},
  {Kipper}, {Tempel}, {Park}, {Gramann}, {Hein{\"a}m{\"a}ki}, {Saar}, \&
  {Einasto}}]{Einasto2018B}
{Einasto}, M., {Deshev}, B., {Lietzen}, H., {et~al.} 2018{\natexlab{a}}, \aap,
  610, A82

\bibitem[{{Einasto} {et~al.}(2020){Einasto}, {Deshev}, {Tenjes},
  {Hein{\"a}m{\"a}ki}, {Tempel}, {Juhan Liivam{\"a}gi}, {Einasto}, {Lietzen},
  {Tuvikene}, \& {Chon}}]{Einasto2020}
{Einasto}, M., {Deshev}, B., {Tenjes}, P., {et~al.} 2020, \aap, 641, A172

\bibitem[{{Einasto} {et~al.}(2018{\natexlab{b}}){Einasto}, {Gramann}, {Park},
  {Kim}, {Deshev}, {Tempel}, {Hein{\"a}m{\"a}ki}, {Lietzen},
  {L{\"a}hteenm{\"a}ki}, {Einasto}, \& {Saar}}]{Einasto2018}
{Einasto}, M., {Gramann}, M., {Park}, C., {et~al.} 2018{\natexlab{b}}, \aap,
  620, A149

\bibitem[{{Einasto} {et~al.}(2015){Einasto}, {Gramann}, {Saar},
  {Liivam{\"a}gi}, {Tempel}, {Nevalainen}, {Hein{\"a}m{\"a}ki}, {Park}, \&
  {Einasto}}]{Einasto2015}
{Einasto}, M., {Gramann}, M., {Saar}, E., {et~al.} 2015, \aap, 580, A69

\bibitem[{{Evans} \& {Bridle}(2009)}]{Evans2009}
{Evans}, A. K.~D. \& {Bridle}, S. 2009, \apj, 695, 1446

\bibitem[{{Fitchett} \& {Webster}(1987)}]{Fitchett1987}
{Fitchett}, M. \& {Webster}, R. 1987, \apj, 317, 653

\bibitem[{{Gal{\'a}rraga-Espinosa}
  {et~al.}(2020{\natexlab{a}}){Gal{\'a}rraga-Espinosa}, {Aghanim}, {Langer},
  {Gouin}, \& {Malavasi}}]{DANY_2020a}
{Gal{\'a}rraga-Espinosa}, D., {Aghanim}, N., {Langer}, M., {Gouin}, C., \&
  {Malavasi}, N. 2020{\natexlab{a}}, \aap, 641, A173

\bibitem[{{Gal{\'a}rraga-Espinosa}
  {et~al.}(2020{\natexlab{b}}){Gal{\'a}rraga-Espinosa}, {Aghanim}, {Langer}, \&
  {Tanimura}}]{DANY_2020b}
{Gal{\'a}rraga-Espinosa}, D., {Aghanim}, N., {Langer}, M., \& {Tanimura}, H.
  2020{\natexlab{b}}, arXiv e-prints, arXiv:2010.15139

\bibitem[{{Gavazzi} {et~al.}(2009){Gavazzi}, {Adami}, {Durret}, {Cuillandre},
  {Ilbert}, {Mazure}, {Pell{\'o}}, \& {Ulmer}}]{Gavazzi2009}
{Gavazzi}, R., {Adami}, C., {Durret}, F., {et~al.} 2009, \aap, 498, L33

\bibitem[{{Giocoli} {et~al.}(2012){Giocoli}, {Tormen}, \&
  {Sheth}}]{Giocoli2012}
{Giocoli}, C., {Tormen}, G., \& {Sheth}, R.~K. 2012, \mnras, 422, 185

\bibitem[{{Gonzalez} {et~al.}(2021){Gonzalez}, {Makler}, {Garc{\'\i}a Lambas},
  {Chalela}, {Pereira}, {Van Waerbeke}, {Shan}, \& {Erben}}]{Gonzalez2021}
{Gonzalez}, E.~J., {Makler}, M., {Garc{\'\i}a Lambas}, D., {et~al.} 2021,
  \mnras, 501, 5239

\bibitem[{{Gouin} {et~al.}(2020){Gouin}, {Aghanim}, {Bonjean}, \&
  {Douspis}}]{Gouin2020}
{Gouin}, C., {Aghanim}, N., {Bonjean}, V., \& {Douspis}, M. 2020, \aap, 635,
  A195

\bibitem[{{Gouin} {et~al.}(2017){Gouin}, {Gavazzi}, {Codis}, {Pichon},
  {Peirani}, \& {Dubois}}]{Gouin2017}
{Gouin}, C., {Gavazzi}, R., {Codis}, S., {et~al.} 2017, \aap, 605, A27

\bibitem[{{Haggar} {et~al.}(2020){Haggar}, {Gray}, {Pearce}, {Knebe}, {Cui},
  {Mostoghiu}, \& {Yepes}}]{Haggar2020}
{Haggar}, R., {Gray}, M.~E., {Pearce}, F.~R., {et~al.} 2020, \mnras, 492, 6074

\bibitem[{{Harvey} {et~al.}(2021){Harvey}, {Robertson}, {Tam}, {Jauzac},
  {Massey}, {Rhodes}, \& {McCarthy}}]{Harvey2021}
{Harvey}, D., {Robertson}, A., {Tam}, S.-I., {et~al.} 2021, \mnras, 500, 2627

\bibitem[{{Hellwing} {et~al.}(2020){Hellwing}, {Cautun}, {van de Weygaert}, \&
  {Jones}}]{Hellwing2020}
{Hellwing}, W.~A., {Cautun}, M., {van de Weygaert}, R., \& {Jones}, B.~T. 2020,
  arXiv e-prints, arXiv:2011.08840

\bibitem[{{Hetznecker} \& {Burkert}(2006)}]{Hetznecker2006}
{Hetznecker}, H. \& {Burkert}, A. 2006, \mnras, 370, 1905

\bibitem[{{Jing} \& {Suto}(2002)}]{Jing2002}
{Jing}, Y.~P. \& {Suto}, Y. 2002, \apj, 574, 538

\bibitem[{{Kent} \& {Gunn}(1982)}]{Kent1982}
{Kent}, S.~M. \& {Gunn}, J.~E. 1982, \aj, 87, 945

\bibitem[{{Klypin} {et~al.}(2016){Klypin}, {Yepes}, {Gottl{\"o}ber}, {Prada},
  \& {He{\ss}}}]{Klypin2016}
{Klypin}, A., {Yepes}, G., {Gottl{\"o}ber}, S., {Prada}, F., \& {He{\ss}}, S.
  2016, \mnras, 457, 4340

\bibitem[{{Kraljic} {et~al.}(2020){Kraljic}, {Pichon}, {Codis}, {Laigle},
  {Dav{\'e}}, {Dubois}, {Hwang}, {Pogosyan}, {Arnouts}, {Devriendt}, {Musso},
  {Peirani}, {Slyz}, \& {Treyer}}]{Kraljic2020}
{Kraljic}, K., {Pichon}, C., {Codis}, S., {et~al.} 2020, \mnras, 491, 4294

\bibitem[{{Kuchner} {et~al.}(2020){Kuchner}, {Arag{\'o}n-Salamanca}, {Pearce},
  {Gray}, {Rost}, {Mu}, {Welker}, {Cui}, {Haggar}, {Laigle}, {Knebe},
  {Kraljic}, {Sarron}, \& {Yepes}}]{Kuchner2020}
{Kuchner}, U., {Arag{\'o}n-Salamanca}, A., {Pearce}, F.~R., {et~al.} 2020,
  \mnras, 494, 5473

\bibitem[{{Kuutma} {et~al.}(2020){Kuutma}, {Poudel}, {Einasto},
  {Hein{\"a}m{\"a}ki}, {Lietzen}, {Tamm}, \& {Tempel}}]{Kuutma2020}
{Kuutma}, T., {Poudel}, A., {Einasto}, M., {et~al.} 2020, \aap, 639, A71

\bibitem[{{Lau} {et~al.}(2021){Lau}, {Hearin}, {Nagai}, \&
  {Cappelluti}}]{Lau2021}
{Lau}, E.~T., {Hearin}, A.~P., {Nagai}, D., \& {Cappelluti}, N. 2021, \mnras,
  500, 1029

\bibitem[{{Lee} {et~al.}(2018){Lee}, {Le Brun}, {Haq}, {Deering}, {King},
  {Applegate}, \& {McCarthy}}]{Lee2018}
{Lee}, B.~E., {Le Brun}, A.~M.~C., {Haq}, M.~E., {et~al.} 2018, \mnras, 479,
  890

\bibitem[{{Lee} {et~al.}(2019){Lee}, {Im}, {Hyun}, {Park}, {Kim}, {Kim}, \&
  {Kim}}]{2019Lee}
{Lee}, S.-K., {Im}, M., {Hyun}, M., {et~al.} 2019, \mnras, 490, 135

\bibitem[{Libeskind {et~al.}(2017)Libeskind, van~de Weygaert, Cautun, Falck,
  Tempel, Abel, Alpaslan, Arag{\'{o}}n-Calvo, Forero-Romero, Gonzalez,
  Gottl{\"{o}}ber, Hahn, Hellwing, Hoffman, Jones, Kitaura, Knebe, Manti,
  Neyrinck, Nuza, Padilla, Platen, Ramachandra, Robotham, Saar, Shandarin,
  Steinmetz, Stoica, Sousbie, \& Yepes}]{Libeskind2017}
Libeskind, N.~I., van~de Weygaert, R., Cautun, M., {et~al.} 2017, Monthly
  Notices of the Royal Astronomical Society, 473, 1195

\bibitem[{{{\L}okas} \& {Mamon}(2003)}]{Lokas2003}
{{\L}okas}, E.~L. \& {Mamon}, G.~A. 2003, \mnras, 343, 401

\bibitem[{{Macci{\`o}} {et~al.}(2007){Macci{\`o}}, {Dutton}, {van den Bosch},
  {Moore}, {Potter}, \& {Stadel}}]{Maccio2007}
{Macci{\`o}}, A.~V., {Dutton}, A.~A., {van den Bosch}, F.~C., {et~al.} 2007,
  \mnras, 378, 55

\bibitem[{{Mahajan} {et~al.}(2018){Mahajan}, {Singh}, \&
  {Shobhana}}]{Mahajan2018}
{Mahajan}, S., {Singh}, A., \& {Shobhana}, D. 2018, \mnras, 478, 4336

\bibitem[{{Malavasi} {et~al.}(2020){Malavasi}, {Aghanim}, {Tanimura},
  {Bonjean}, \& {Douspis}}]{Malavasi2020}
{Malavasi}, N., {Aghanim}, N., {Tanimura}, H., {Bonjean}, V., \& {Douspis}, M.
  2020, \aap, 634, A30

\bibitem[{{More} {et~al.}(2015){More}, {Diemer}, \& {Kravtsov}}]{More2015}
{More}, S., {Diemer}, B., \& {Kravtsov}, A.~V. 2015, \apj, 810, 36

\bibitem[{{Morinaga} \& {Ishiyama}(2020)}]{Morinaga2020}
{Morinaga}, Y. \& {Ishiyama}, T. 2020, \mnras, 495, 502

\bibitem[{{Mostoghiu} {et~al.}(2019){Mostoghiu}, {Knebe}, {Cui}, {Pearce},
  {Yepes}, {Power}, {Dave}, \& {Arth}}]{Mostoghiu2019}
{Mostoghiu}, R., {Knebe}, A., {Cui}, W., {et~al.} 2019, \mnras, 483, 3390

\bibitem[{{Munari} {et~al.}(2014){Munari}, {Biviano}, \& {Mamon}}]{Munari2014}
{Munari}, E., {Biviano}, A., \& {Mamon}, G.~A. 2014, \aap, 566, A68

\bibitem[{{Musso} {et~al.}(2018){Musso}, {Cadiou}, {Pichon}, {Codis},
  {Kraljic}, \& {Dubois}}]{Musso2018}
{Musso}, M., {Cadiou}, C., {Pichon}, C., {et~al.} 2018, \mnras, 476, 4877

\bibitem[{{Nelson} {et~al.}(2019){Nelson}, {Springel}, {Pillepich},
  {Rodriguez-Gomez}, {Torrey}, {Genel}, {Vogelsberger}, {Pakmor}, {Marinacci},
  {Weinberger}, {Kelley}, {Lovell}, {Diemer}, \& {Hernquist}}]{ILLUSTRIS_TNG}
{Nelson}, D., {Springel}, V., {Pillepich}, A., {et~al.} 2019, Computational
  Astrophysics and Cosmology, 6, 2

\bibitem[{{Neto} {et~al.}(2007){Neto}, {Gao}, {Bett}, {Cole}, {Navarro},
  {Frenk}, {White}, {Springel}, \& {Jenkins}}]{Neto2007}
{Neto}, A.~F., {Gao}, L., {Bett}, P., {et~al.} 2007, \mnras, 381, 1450

\bibitem[{{Neumann} {et~al.}(2003){Neumann}, {Lumb}, {Pratt}, \&
  {Briel}}]{Neumann2003}
{Neumann}, D.~M., {Lumb}, D.~H., {Pratt}, G.~W., \& {Briel}, U.~G. 2003, \aap,
  400, 811

\bibitem[{{Oguri} {et~al.}(2010){Oguri}, {Takada}, {Okabe}, \&
  {Smith}}]{Oguri2010}
{Oguri}, M., {Takada}, M., {Okabe}, N., \& {Smith}, G.~P. 2010, \mnras, 405,
  2215

\bibitem[{{Okabe} {et~al.}(2018){Okabe}, {Nishimichi}, {Oguri}, {Peirani},
  {Kitayama}, {Sasaki}, \& {Suto}}]{Okabe2018}
{Okabe}, T., {Nishimichi}, T., {Oguri}, M., {et~al.} 2018, \mnras, 478, 1141

\bibitem[{{Okabe} {et~al.}(2020{\natexlab{a}}){Okabe}, {Nishimichi}, {Oguri},
  {Peirani}, {Kitayama}, {Sasaki}, {Suto}, {Pichon}, \&
  {Dubois}}]{Okabe2020_SIMU}
{Okabe}, T., {Nishimichi}, T., {Oguri}, M., {et~al.} 2020{\natexlab{a}},
  \mnras, 491, 2268

\bibitem[{{Okabe} {et~al.}(2020{\natexlab{b}}){Okabe}, {Oguri}, {Peirani},
  {Suto}, {Dubois}, {Pichon}, {Kitayama}, {Sasaki}, \&
  {Nishimichi}}]{Okabe2020_OBS}
{Okabe}, T., {Oguri}, M., {Peirani}, S., {et~al.} 2020{\natexlab{b}}, \mnras,
  496, 2591

\bibitem[{{Paranjape} {et~al.}(2018){Paranjape}, {Hahn}, \&
  {Sheth}}]{Paranjape2018}
{Paranjape}, A., {Hahn}, O., \& {Sheth}, R.~K. 2018, \mnras, 476, 3631

\bibitem[{{Peter} {et~al.}(2013){Peter}, {Rocha}, {Bullock}, \&
  {Kaplinghat}}]{Peter2013}
{Peter}, A. H.~G., {Rocha}, M., {Bullock}, J.~S., \& {Kaplinghat}, M. 2013,
  \mnras, 430, 105

\bibitem[{{Pichon} {et~al.}(2010){Pichon}, {Gay}, {Pogosyan}, {Prunet},
  {Sousbie}, {Colombi}, {Slyz}, \& {Devriendt}}]{Pichon2010}
{Pichon}, C., {Gay}, C., {Pogosyan}, D., {et~al.} 2010, in American Institute
  of Physics Conference Series, Vol. 1241, American Institute of Physics
  Conference Series, ed. J.-M. {Alimi} \& A.~{Fu{\"o}zfa}, 1108--1117

\bibitem[{{Planck Collaboration} {et~al.}(2016){Planck Collaboration}, {Ade},
  {Aghanim}, {Arnaud}, {Ashdown}, {Aumont}, {Baccigalupi}, {Banday},
  {Barreiro}, {Bartlett}, {Bartolo}, {Battaner}, {Battye}, {Benabed},
  {Beno{\^\i}t}, {Benoit-L{\'e}vy}, {Bernard}, {Bersanelli}, {Bielewicz},
  {Bock}, {Bonaldi}, {Bonavera}, {Bond}, {Borrill}, {Bouchet}, {Bucher},
  {Burigana}, {Butler}, {Calabrese}, {Cardoso}, {Catalano}, {Challinor},
  {Chamballu}, {Chary}, {Chiang}, {Christensen}, {Church}, {Clements},
  {Colombi}, {Colombo}, {Combet}, {Comis}, {Couchot}, {Coulais}, {Crill},
  {Curto}, {Cuttaia}, {Danese}, {Davies}, {Davis}, {de Bernardis}, {de Rosa},
  {de Zotti}, {Delabrouille}, {D{\'e}sert}, {Diego}, {Dolag}, {Dole},
  {Donzelli}, {Dor{\'e}}, {Douspis}, {Ducout}, {Dupac}, {Efstathiou}, {Elsner},
  {En{\ss}lin}, {Eriksen}, {Falgarone}, {Fergusson}, {Finelli}, {Forni},
  {Frailis}, {Fraisse}, {Franceschi}, {Frejsel}, {Galeotta}, {Galli}, {Ganga},
  {Giard}, {Giraud-H{\'e}raud}, {Gjerl{\o}w}, {Gonz{\'a}lez-Nuevo},
  {G{\'o}rski}, {Gratton}, {Gregorio}, {Gruppuso}, {Gudmundsson}, {Hansen},
  {Hanson}, {Harrison}, {Henrot-Versill{\'e}}, {Hern{\'a}ndez-Monteagudo},
  {Herranz}, {Hildebrandt}, {Hivon}, {Hobson}, {Holmes}, {Hornstrup}, {Hovest},
  {Huffenberger}, {Hurier}, {Jaffe}, {Jaffe}, {Jones}, {Juvela},
  {Keih{\"a}nen}, {Keskitalo}, {Kisner}, {Kneissl}, {Knoche}, {Kunz},
  {Kurki-Suonio}, {Lagache}, {L{\"a}hteenm{\"a}ki}, {Lamarre}, {Lasenby},
  {Lattanzi}, {Lawrence}, {Leonardi}, {Lesgourgues}, {Levrier}, {Liguori},
  {Lilje}, {Linden-V{\o}rnle}, {L{\'o}pez-Caniego}, {Lubin},
  {Mac{\'\i}as-P{\'e}rez}, {Maggio}, {Maino}, {Mand olesi}, {Mangilli},
  {Maris}, {Martin}, {Mart{\'\i}nez-Gonz{\'a}lez}, {Masi}, {Matarrese},
  {McGehee}, {Meinhold}, {Melchiorri}, {Melin}, {Mendes}, {Mennella},
  {Migliaccio}, {Mitra}, {Miville-Desch{\^e}nes}, {Moneti}, {Montier},
  {Morgante}, {Mortlock}, {Moss}, {Munshi}, {Murphy}, {Naselsky}, {Nati},
  {Natoli}, {Netterfield}, {N{\o}rgaard-Nielsen}, {Noviello}, {Novikov},
  {Novikov}, {Oxborrow}, {Paci}, {Pagano}, {Pajot}, {Paoletti}, {Partridge},
  {Pasian}, {Patanchon}, {Pearson}, {Perdereau}, {Perotto}, {Perrotta},
  {Pettorino}, {Piacentini}, {Piat}, {Pierpaoli}, {Pietrobon}, {Plaszczynski},
  {Pointecouteau}, {Polenta}, {Popa}, {Pratt}, {Pr{\'e}zeau}, {Prunet},
  {Puget}, {Rachen}, {Rebolo}, {Reinecke}, {Remazeilles}, {Renault}, {Renzi},
  {Ristorcelli}, {Rocha}, {Roman}, {Rosset}, {Rossetti}, {Roudier},
  {Rubi{\~n}o-Mart{\'\i}n}, {Rusholme}, {Sandri}, {Santos}, {Savelainen},
  {Savini}, {Scott}, {Seiffert}, {Shellard}, {Spencer}, {Stolyarov}, {Stompor},
  {Sudiwala}, {Sunyaev}, {Sutton}, {Suur-Uski}, {Sygnet}, {Tauber}, {Terenzi},
  {Toffolatti}, {Tomasi}, {Tristram}, {Tucci}, {Tuovinen}, {T{\"u}rler},
  {Umana}, {Valenziano}, {Valiviita}, {Van Tent}, {Vielva}, {Villa}, {Wade},
  {Wandelt}, {Wehus}, {Weller}, {White}, {Yvon}, {Zacchei}, \&
  {Zonca}}]{Planck2016}
{Planck Collaboration}, {Ade}, P.~A.~R., {Aghanim}, N., {et~al.} 2016, \aap,
  594, A24

\bibitem[{{Poole} {et~al.}(2017){Poole}, {Mutch}, {Croton}, \&
  {Wyithe}}]{Poole2017}
{Poole}, G.~B., {Mutch}, S.~J., {Croton}, D.~J., \& {Wyithe}, S. 2017, \mnras,
  472, 3659

\bibitem[{{Poudel} {et~al.}(2017){Poudel}, {Hein{\"a}m{\"a}ki}, {Tempel},
  {Einasto}, {Lietzen}, \& {Nurmi}}]{Poudel2017}
{Poudel}, A., {Hein{\"a}m{\"a}ki}, P., {Tempel}, E., {et~al.} 2017, \aap, 597,
  A86

\bibitem[{{Power} {et~al.}(2012){Power}, {Knebe}, \& {Knollmann}}]{Power2012}
{Power}, C., {Knebe}, A., \& {Knollmann}, S.~R. 2012, \mnras, 419, 1576

\bibitem[{{Ramakrishnan} {et~al.}(2019){Ramakrishnan}, {Paranjape}, {Hahn}, \&
  {Sheth}}]{Ramakrishnan2019}
{Ramakrishnan}, S., {Paranjape}, A., {Hahn}, O., \& {Sheth}, R.~K. 2019,
  \mnras, 489, 2977

\bibitem[{{Rodriguez-Gomez} {et~al.}(2015){Rodriguez-Gomez}, {Genel},
  {Vogelsberger}, {Sijacki}, {Pillepich}, {Sales}, {Torrey}, {Snyder},
  {Nelson}, {Springel}, {Ma}, \& {Hernquist}}]{Rodriguez2015}
{Rodriguez-Gomez}, V., {Genel}, S., {Vogelsberger}, M., {et~al.} 2015, \mnras,
  449, 49

\bibitem[{{Rodr{\'\i}guez-Puebla} {et~al.}(2016){Rodr{\'\i}guez-Puebla},
  {Behroozi}, {Primack}, {Klypin}, {Lee}, \& {Hellinger}}]{Rodriguez2016}
{Rodr{\'\i}guez-Puebla}, A., {Behroozi}, P., {Primack}, J., {et~al.} 2016,
  \mnras, 462, 893

\bibitem[{{Rossetti} {et~al.}(2013){Rossetti}, {Eckert}, {De Grandi},
  {Gastaldello}, {Ghizzardi}, {Roediger}, \& {Molendi}}]{Rossetti2013}
{Rossetti}, M., {Eckert}, D., {De Grandi}, S., {et~al.} 2013, \aap, 556, A44

\bibitem[{{Rossetti} {et~al.}(2016){Rossetti}, {Gastaldello}, {Ferioli},
  {Bersanelli}, {De Grandi}, {Eckert}, {Ghizzardi}, {Maino}, \&
  {Molendi}}]{Rossetti2016}
{Rossetti}, M., {Gastaldello}, F., {Ferioli}, G., {et~al.} 2016, \mnras, 457,
  4515

\bibitem[{{Rykoff} {et~al.}(2016){Rykoff}, {Rozo}, {Hollowood}, {Bermeo-Hernand
  ez}, {Jeltema}, {Mayers}, {Romer}, {Rooney}, {Saro}, {Vergara Cervantes},
  {Wechsler}, {Wilcox}, {Abbott}, {Abdalla}, {Allam}, {Annis},
  {Benoit-L{\'e}vy}, {Bernstein}, {Bertin}, {Brooks}, {Burke}, {Capozzi},
  {Carnero Rosell}, {Carrasco Kind}, {Castander}, {Childress}, {Collins},
  {Cunha}, {D'Andrea}, {da Costa}, {Davis}, {Desai}, {Diehl}, {Dietrich},
  {Doel}, {Evrard}, {Finley}, {Flaugher}, {Fosalba}, {Frieman}, {Glazebrook},
  {Goldstein}, {Gruen}, {Gruendl}, {Gutierrez}, {Hilton}, {Honscheid}, {Hoyle},
  {James}, {Kay}, {Kuehn}, {Kuropatkin}, {Lahav}, {Lewis}, {Lidman}, {Lima},
  {Maia}, {Mann}, {Marshall}, {Martini}, {Melchior}, {Miller}, {Miquel},
  {Mohr}, {Nichol}, {Nord}, {Ogando}, {Plazas}, {Reil}, {Sahl{\'e}n},
  {Sanchez}, {Santiago}, {Scarpine}, {Schubnell}, {Sevilla-Noarbe}, {Smith},
  {Soares-Santos}, {Sobreira}, {Stott}, {Suchyta}, {Swanson}, {Tarle},
  {Thomas}, {Tucker}, {Uddin}, {Viana}, {Vikram}, {Walker}, {Zhang}, \& {DES
  Collaboration}}]{Redmapper}
{Rykoff}, E.~S., {Rozo}, E., {Hollowood}, D., {et~al.} 2016, \apjs, 224, 1

\bibitem[{{Sarron} {et~al.}(2019){Sarron}, {Adami}, {Durret}, \&
  {Laigle}}]{Sarron2019}
{Sarron}, F., {Adami}, C., {Durret}, F., \& {Laigle}, C. 2019, \aap, 632, A49

\bibitem[{{Sereno} {et~al.}(2018){Sereno}, {Umetsu}, {Ettori}, {Sayers},
  {Chiu}, {Meneghetti}, {Vega-Ferrero}, \& {Zitrin}}]{Sereno2018}
{Sereno}, M., {Umetsu}, K., {Ettori}, S., {et~al.} 2018, \apjl, 860, L4

\bibitem[{{Shim} {et~al.}(2020){Shim}, {Codis}, {Pichon}, {Pogosyan}, \&
  {Cadiou}}]{Shim2020}
{Shim}, J., {Codis}, S., {Pichon}, C., {Pogosyan}, D., \& {Cadiou}, C. 2020,
  arXiv e-prints, arXiv:2011.04321

\bibitem[{{Shin} {et~al.}(2018){Shin}, {Clampitt}, {Jain}, {Bernstein}, {Neil},
  {Rozo}, \& {Rykoff}}]{Shin2018}
{Shin}, T.-h., {Clampitt}, J., {Jain}, B., {et~al.} 2018, \mnras, 475, 2421

\bibitem[{{Spergel} \& {Steinhardt}(2000)}]{Spergel2000}
{Spergel}, D.~N. \& {Steinhardt}, P.~J. 2000, \prl, 84, 3760

\bibitem[{{Springel}(2010)}]{AREPO}
{Springel}, V. 2010, \mnras, 401, 791

\bibitem[{{Springel} {et~al.}(2001){Springel}, {White}, {Tormen}, \&
  {Kauffmann}}]{subfind}
{Springel}, V., {White}, S. D.~M., {Tormen}, G., \& {Kauffmann}, G. 2001,
  \mnras, 328, 726

\bibitem[{{Suto} {et~al.}(2016){Suto}, {Kitayama}, {Nishimichi}, {Sasaki}, \&
  {Suto}}]{Suto2016}
{Suto}, D., {Kitayama}, T., {Nishimichi}, T., {Sasaki}, S., \& {Suto}, Y. 2016,
  \pasj, 68, 97

\bibitem[{{The CHEX-MATE Collaboration} {et~al.}(2020){The CHEX-MATE
  Collaboration}, {:}, {Arnaud}, {Ettori}, {Pratt}, {Rossetti}, {Eckert},
  {Gastaldello}, {Gavazzi}, {Kay}, {Lovisari}, {Maughan}, {Pointecouteau},
  {Sereno}, {Bartalucci}, {Bonafede}, {Bourdin}, {Cassano}, {Duffy}, {Iqbal},
  {Maurogordato}, {Rasia}, {Sayers}, {Andrade-Santos}, {Aussel}, {Barnes},
  {Barrena}, {Borgani}, {Burkutean}, {Clerc}, {Corasaniti}, {Cuillandre}, {De
  Grandi}, {De Petris}, {Dolag}, {Donahue}, {Ferragamo}, {Gaspari},
  {Ghizzardi}, {Gitti}, {Haines}, {Jauzac}, {Johnston-Hollitt}, {Jones},
  {K{\'e}ruzor{\'e}}, {Le Brun}, {Mayet}, {Mazzotta}, {Melin}, {Molendi},
  {Nonino}, {Okabe}, {Paltani}, {Perotto}, {Pires}, {Radovich},
  {Rubino-Martin}, {Salvati}, {Saro}, {Sartoris}, {Schellenberger},
  {Streblyanska}, {Tarrio}, {Tozzi}, {Umetsu}, {van der Burg}, {Vazza},
  {Venturi}, {Yepes}, \& {Zarattini}}]{Heritage2020}
{The CHEX-MATE Collaboration}, {:}, {Arnaud}, M., {et~al.} 2020, arXiv
  e-prints, arXiv:2010.11972

\bibitem[{{Thierbach} {et~al.}(2003){Thierbach}, {Klein}, \&
  {Wielebinski}}]{Thierbach2003}
{Thierbach}, M., {Klein}, U., \& {Wielebinski}, R. 2003, \aap, 397, 53

\bibitem[{{Vall{\'e}s-P{\'e}rez} {et~al.}(2020){Vall{\'e}s-P{\'e}rez},
  {Planelles}, \& {Quilis}}]{Valles2020}
{Vall{\'e}s-P{\'e}rez}, D., {Planelles}, S., \& {Quilis}, V. 2020, \mnras, 499,
  2303

\bibitem[{{van den Bosch}(2002)}]{vandenBosch2002}
{van den Bosch}, F.~C. 2002, \mnras, 331, 98

\bibitem[{{Vega-Ferrero} {et~al.}(2017){Vega-Ferrero}, {Yepes}, \&
  {Gottl{\"o}ber}}]{Vega2017}
{Vega-Ferrero}, J., {Yepes}, G., \& {Gottl{\"o}ber}, S. 2017, \mnras, 467, 3226

\bibitem[{{Velliscig} {et~al.}(2015){Velliscig}, {Cacciato}, {Schaye}, {Crain},
  {Bower}, {van Daalen}, {Dalla Vecchia}, {Frenk}, {Furlong}, {McCarthy},
  {Schaller}, \& {Theuns}}]{Velliscig2015}
{Velliscig}, M., {Cacciato}, M., {Schaye}, J., {et~al.} 2015, \mnras, 453, 721

\bibitem[{{Wechsler} {et~al.}(2002){Wechsler}, {Bullock}, {Primack},
  {Kravtsov}, \& {Dekel}}]{Wechsler2002}
{Wechsler}, R.~H., {Bullock}, J.~S., {Primack}, J.~R., {Kravtsov}, A.~V., \&
  {Dekel}, A. 2002, \apj, 568, 52

\bibitem[{{Wong} \& {Taylor}(2012)}]{Wong2012}
{Wong}, A. W.~C. \& {Taylor}, J.~E. 2012, \apj, 757, 102

\bibitem[{{Yoshida} {et~al.}(2000){Yoshida}, {Springel}, {White}, \&
  {Tormen}}]{Yoshida2000}
{Yoshida}, N., {Springel}, V., {White}, S. D.~M., \& {Tormen}, G. 2000, \apjl,
  544, L87

\bibitem[{{Yuan} \& {Han}(2020)}]{Yuan2020}
{Yuan}, Z.~S. \& {Han}, J.~L. 2020, \mnras, 497, 5485

\bibitem[{{Zel'Dovich}(1970)}]{Zeldovich1970}
{Zel'Dovich}, Y.~B. 1970, \aap, 500, 13

\end{thebibliography}

\end{document}